\documentclass[lettersize,journal]{IEEEtran}
\usepackage{amsmath,amsfonts}
\usepackage{algorithmic}
\usepackage{algorithm}
\usepackage{array}
\usepackage{amssymb}
\usepackage[caption=false,font=normalsize,labelfont=sf,textfont=sf]{subfig}
\usepackage{textcomp}
\usepackage{stfloats}
\usepackage{url}
\usepackage{verbatim}
\usepackage{graphicx}
\usepackage{cite}
\usepackage{xcolor}
\usepackage{multirow}
\usepackage{booktabs}

\begin{document}

\title{See the Speaker: Crafting High-Resolution Talking Faces from Speech with Prior Guidance and Region Refinement}

\author{
Jinting Wang, 
Jun Wang, 
Hei Victor Cheng\thanks{Jinting Wang and Li Liu are with the Hong Kong University of Science and Technology (Guangzhou) (jwang644@connect.hkust-gz.edu.cn, avrillliu@hkust-gz.edu.cn). Jun Wang is with  Speech and Acoustic Laboratory, Joy Future Academy, Jingdong Corporation (wangjun.judy@jd.com). Hei Victor Cheng is with Aarhus University (hvc@ece.au.dk).}, 
Li Liu\textsuperscript{*},~\IEEEmembership{Senior Member,~IEEE}\thanks{*Corresponding Author: avrillliu@hkust-gz.edu.cn.}
}



\markboth{Journal of \LaTeX\ Class Files,~Vol.~14, No.~8, August~2021}%
{Shell \MakeLowercase{\textit{et al.}}: A Sample Article Using IEEEtran.cls for IEEE Journals}


\maketitle

\begin{abstract} 
Unlike existing methods that rely on source images as appearance references and use source speech to generate motion, this work proposes a novel approach that directly extracts information from the speech, addressing key challenges in speech-to-talking face.
Specifically, we first employ a speech-to-face portrait generation stage, utilizing a speech-conditioned diffusion model combined with statistical facial prior and a sample-adaptive weighting module to achieve high-quality portrait generation. In the subsequent speech-driven talking face generation stage, we embed expressive dynamics such as lip movement, facial expressions, and eye movements into the latent space of the diffusion model and further optimize lip synchronization using a region-enhancement module. To generate high-resolution outputs, we integrate a pre-trained Transformer-based discrete codebook with an image rendering network, enhancing video frame details in an end-to-end manner. Experimental results demonstrate that our method outperforms existing approaches on the HDTF, VoxCeleb, and AVSpeech datasets. Notably, this is the first method capable of generating high-resolution, high-quality talking face videos exclusively from a single speech input.

\end{abstract}

\begin{IEEEkeywords}
Talking face generation, speech-to-portrait, high-resolution, diffusion model, prior knowledge, lip refinement, latent motion representation, discrete codebook.
\end{IEEEkeywords}

\section{Introduction}
\IEEEPARstart{A}{udio}-driven talking face generation aims to animate a target portrait image to create realistic talking videos given a driving audio speech. 
This technique finds wide application in various practical scenarios, including high-quality film and animation production, virtual assistants, interactive educational content creation, and realistic character animation. 

Recently, significant advancements have been made in this field with the development of generative models.
Existing talking face generation methods mainly focus on creating animated videos from a reference portrait \cite{ye2024mimictalk,ye2024real3d,wang2024high, liu2024anitalker,xu2024hallo}. Still, there is a dilemma: users are concerned about privacy breaches when using real portrait images \cite{xu2024facechain}. FaceChain \cite{xu2024facechain} made the first attempt to liberate the source face and directly infer the synchronized portrait using disentangled identity features from speech. However, the generated virtual face fails to preserve identity consistency. 
Additionally, to achieve realistic talking faces, some methods employ explicit motion representation such as landmarks coefficients \cite{wei2024aniportrait,zhong2023identity}, 3D Morphable Models (3DMM) \cite{ma2023dreamtalk,ye2024real3d,danvevcek2022emoca}, or blendshapes \cite{peng2024synctalk,chen2023improving,peng2023emotalk}, to animate facial dynamics. However, the construction of geometric structure is generally estimated from source images. The initial state of face image has a certain impact on the generation which results in generated faces that seem rigid and unconvincing.
The other option is to model motion features with implicit latent space. For example, VASA-1 \cite{xu2024vasa} and Anitalker \cite{liu2024anitalker} predict motion latent probabilistic distribution with diffusion model conditioned on the audio speech and other input signals, which represent expressive facial features and natural head movements in a joint manner for lifelike talking face. However, the other expressive dynamics in holistic motion representation may damage the lip movement consistency. There remains a gap between the generated animations and the genuine human movement patterns.

\begin{figure}[t]
    \centering
    \includegraphics[width=\columnwidth]{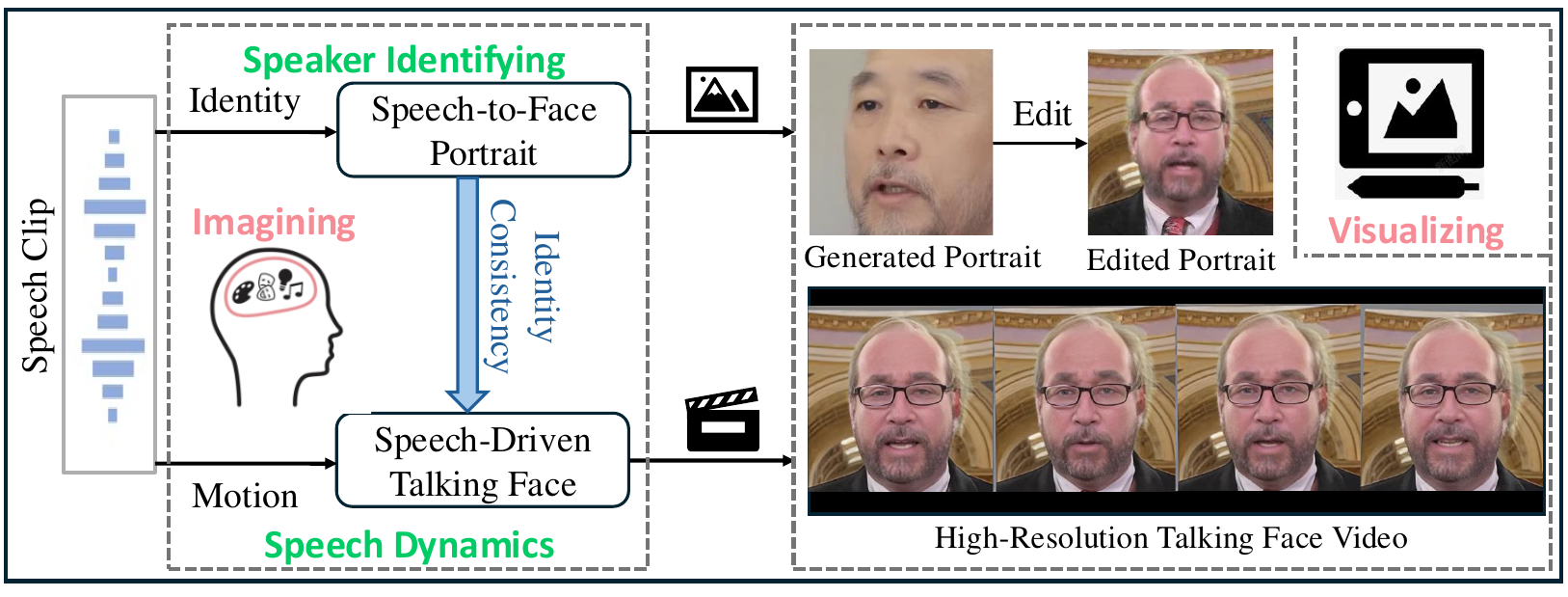}
    \caption{Our framework enables high-resolution talking face video generation from a single audio speech. Firstly, identity information is disentangled to synthesize a speaker's face portrait, followed by the generation of talking videos that align with the decoupled motion cues, all while maintaining identity consistency throughout the video. Notably, for aesthetic purposes and to ensure a fair comparison, we edit the generated face portraits by adding audio-unrelated attributes, such as hair, clothing, and background, etc.}
    \label{fig:overview}
\end{figure}

Video resolution constitutes a critical factor for interactive applications. Existing advances in image and video diffusion have demonstrated significant progress in resolution enhancement through cascaded frameworks, wherein each subsequent latent diffusion model (LDM) is conditioned on the output of the preceding one \cite{ho2022cascaded,ling2024posetalk}. Despite their effectiveness, such approaches introduce additional modules into the pipeline and substantially increase inference overhead. An end-to-end design, by contrast, is a more desirable property for high-resolution talking face generation, both in terms of practical utility and conceptual elegance.

Given the limitations of existing methods, this work develops an effective pipeline for high-resolution talking face video generation from a single audio input. This mirrors an intuitive process, as people often analyze the speech and then mentally visualize the corresponding video clip when listening. As illustrated in Fig. \ref{fig:overview}, we mimic this process by achieving speech-to-portrait generation (S2P) and speech-driven talking face generation (S2TF) progressively, using disentangled information extracted from speech.

\textbf{Firstly, our approach enables high-quality S2P for a diverse range of speakers, capturing real-world scenarios.} Although previous studies have explored the relationship between human speech and facial structures, demonstrating the feasibility of S2P \cite{bull1983voice, deaton2010understanding, schweinberger2014speaker}, the task remains challenging due to the inherent diversity of human faces and the variability in speaking styles. To address this, we propose a speech-conditioned latent diffusion model (LDM) that functions as a personalized portrait generator, guided by a statistical face prior (i.e., general facial features), which is based on the idea that a human face can be decomposed into both general features and personalized characteristics. Additionally, we enhance the speaker-specific portrait variation in the speech by incorporating a sample-adaptive weighted module, which dynamically adjusts the importance of the face prior to better capturing individual differences.

\textbf{Secondly, we address the challenge of generating natural and consistent talking videos.} To better capture expressive dynamics, we incorporate a wide range of motion patterns, including lip movements, facial expressions, eye gaze, and blinking, as latent variables within the latent space of a diffusion model. To prevent the interference of non-lip dynamics on lip movements, we introduce a region enhancement module, which enhances the consistency of lip motion. 

\textbf{Thirdly, we focus on achieving high-resolution video generation.} Discrete prior representations with learned codebook have proven effective for image restoration \cite{razavi2019generating,zhou2022towards}. Unlike prior works that rely on cascaded frameworks, we extend a discrete codebook \cite{zhou2022towards} into the image rendering network, enhancing the quality of generated video frames in an end-to-end manner. By incorporating a high-quality decoder, we ensure smooth transitions in the predicted code sequences, resulting in videos with high-resolution details.

To evaluate the effectiveness of our proposed method, we conducted comprehensive experiments on publicly available datasets, including HDTF \cite{zhang2021flow}, VoxCeleb \cite{chung2018voxceleb2}, and AVSpeech \cite{ephrat2018looking}. To the best of our knowledge, our approach is the first to achieve high-resolution and high-quality talking face generation using only a single audio speech input.





 \begin{figure*}[htbp]
    \centering
    \includegraphics[width=0.8\linewidth]{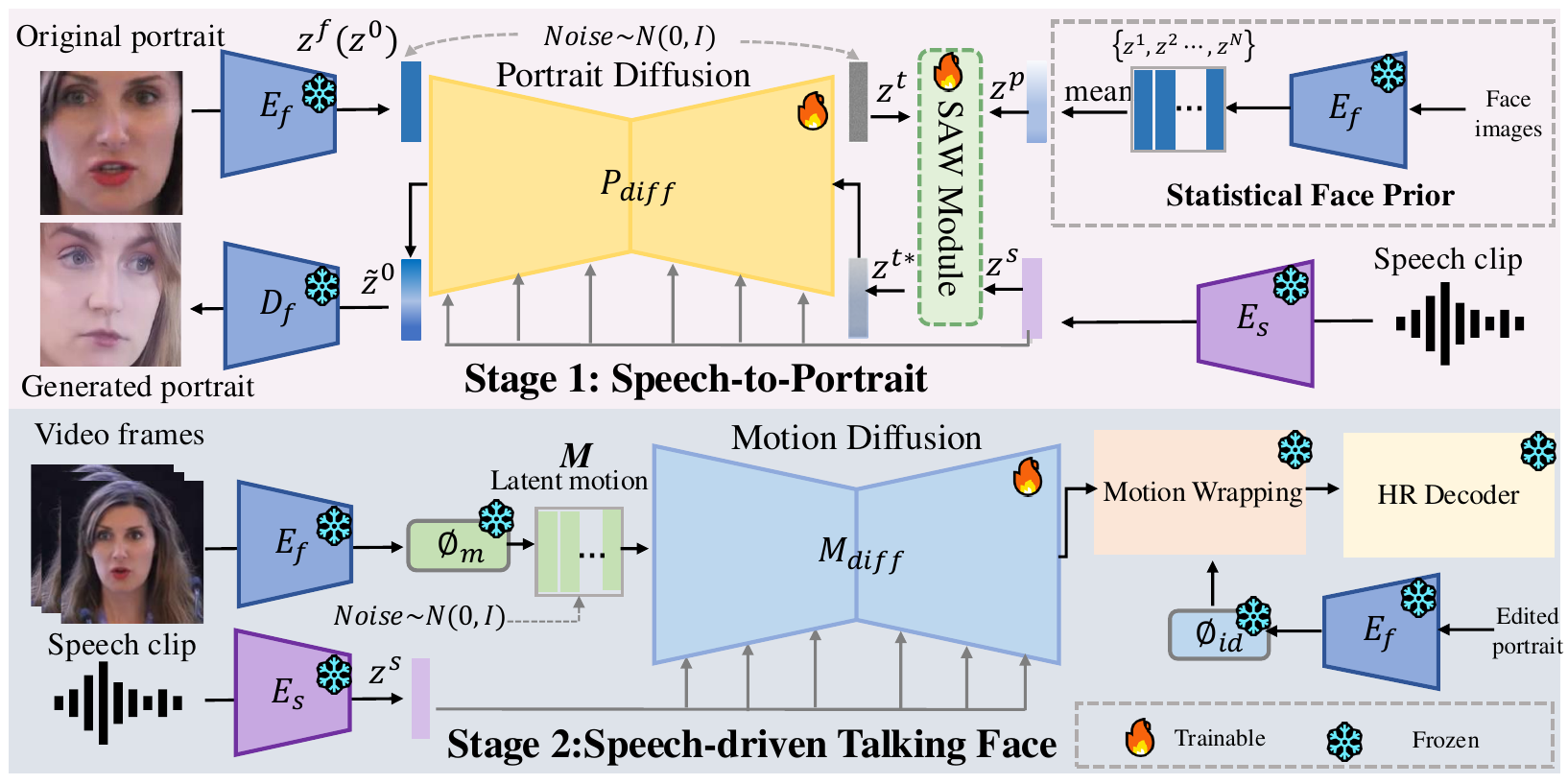}
    \caption{Overview of the proposed two-stage high-resolution talking face generation framework: (1) \textbf{Stage 1: Speech-Conditioned Portrait Generation with Face Prior Guidance (SCFP)}. In this stage, portrait diffusion $P_{diff}$ is trained to capture the personalized speech-portrait correlation using statistical face prior guidance. To emphasize the individual variance conditioned on the speech, we design a Sample-Adaptive Weighted (SAW) module that adaptively adjusts the face prior weight on the noise input. (2) \textbf{Stage 2: High-Resolution Talking Face Synthesis with Holistic Motion and Lip Region Refinement (HRTF).} Based on the speech condition, we develop a motion diffusion $M_{diff}$, to capture the holistic motion representation, including both facial dynamics and head movement, in the latent space. Subsequently, a motion wrapping module and a high-resolution decoder render the learned motion into high-resolution talking face videos, preserving both the static and dynamic visual attributes of the target identity.}
    \label{fig:framework}
\end{figure*}

\section{Related Work}
\subsection{Speech-to-Portrait Generation}
Audio-visual cross-modal learning, particularly S2P, has gained significant attention in recent years. Many existing methods in this field employed GAN-based frameworks. For example, Wav2Pix \cite{duarte2019wav2pix} proposed a speech-conditioned portrait generation framework, which was relatively simplistic and overlooked the preservation of identity information during the generation process.
To address the preservation of identity information, Wen \textit{et al.} \cite{wen2019face} and Fang \textit{et al.} \cite{fang2022facial} designed networks capable of generating portraits from speech by matching the identities of the generated portrait with those of the speakers. 
While explicitly modeling the identity relevance between speech and face portrait modalities was beneficial for ensuring the authenticity of generated images, it had limitations when attempting to generate portraits of different identities. On the other hand, Choi \textit{et al.} \cite{choi2019inference} proposed a two-stage framework for more flexibility in generating face portraits with different identities.
GAN-based methods were often difficult to
train and easy to collapse without careful design, collapsing without carefully selected hyperparameters and regularizers \cite{dhariwal2021diffusion}.
In \cite{oh2019speech2face} and \cite{bai2022speech}, a CNN-based method was proposed for random identity face generation.
The above methods for S2P have shown some progress, but they still have limitations, particularly in terms of generation quality. Inspired by the good performance of LDMs, Kato \textit{et al.} utilized two LDMs for S2P and image quality enhancement. FaceChain \cite{xu2024facechain}  leveraged the LDM for realistic faces generation.

However, LDM is sensitive to the input noise, which would result in generation variances with the same speech condition and poor condition consistency.
In this work, we propose a S2P network that 
introduces a statistical face prior to the input noise to alleviate the output diversity and improve the condition consistency.

\subsection{Audio-driven Talking Face Generation}
Existing audio-driven talking head generation techniques can be broadly categorized into two primary approaches: generating talking head videos with or without an intermediate representation. The use of an intermediate representation allows for the direct or indirect incorporation of additional control signals, which can guide the video generation process. For example, Vividtalk \cite{sun2023vividtalk} proposes synthesizing head motion and facial expressions, which are then used to construct a 3D facial mesh. This mesh serves as an intermediate representation to steer the generation of the final video frames. Similarly, Sadtalker \cite{zhang2023sadtalker} and Real3d-portrait \cite{2024real3d} adopt a 3D Morphable Model (3DMM) as an intermediate representation to produce talking head videos. Additionally, Dreamtalk \cite{ma2023dreamtalk} integrates diffusion models to generate coefficients for the 3DMM, further enhancing the control over the generated video content. SyncTalk \cite{peng2024synctalk} utilizes a 3D facial blendshape model to capture accurate facial expressions, combined with a Face-Sync Controller to align lip movements with speech.
Current works like AniPortrait \cite{wei2024aniportrait} also generate talking head videos by first extracting the 3D facial mesh and head pose from the audio, and then synthesizing video frames conditioned on these pose parameters using diffusion models. However, a common challenge across these techniques is the limited ability of the 3D mesh to capture nuanced details, which constrains the dynamic range and authenticity of the synthesized video sequences. 
In contrast, methods that do not rely on intermediate control signals for audio-driven video generation tend to exhibit higher naturalness and better identity preservation, maintaining consistency with the original image.
For example, EMO \cite{tian2025emo} takes a direct audio-to-video synthesis approach, generating expressive portrait videos with an audio2video diffusion model under weak supervision, without the need for intermediate 3D models or facial landmarks. Hallo \cite{xu2024hallo} introduces a hierarchical audio-driven visual synthesis approach that uses bounding box masks for the lips, expressions, and head. This technique allows for refined control over the diversity of facial expressions and pose variations. VASA-1 \cite{xu2024vasa} and Anitalker \cite{liu2024anitalker} integrate nuanced facial expressions and universal motion representations, resulting in lifelike and synchronous animations.

However, methods that bypass intermediate representations and model holistic motion features in the latent space of diffusion models may suffer from inconsistencies in lip movement synchronization. This can impact the overall coherence and naturalness of lip movements, ultimately affecting the video’s authenticity.

\subsection{High-Resolution Image/Video Generation}
Recent advances have significantly enhanced the generation of high-resolution image/video generation.
Cascade models have been explored in high-resolution image generation \cite{ho2022cascaded, ho2022imagen, podell2023sdxl}, which comprises a pipeline of multiple models that generate images of increasing resolution. 
For example, Cascaded Diffusion Models \cite{ho2022cascaded} cascades several diffusion based super-resolution models behind a diffusion model, but its application remains capped at $256^2$ resolution images. SDXL \cite{podell2023sdxl} introduces a refinement model to achieve $1024^2$ resolution images using a post-hoc image-to-image technique. Posetalk \cite{ling2024posetalk}, Blattmann \textit{et al.} \cite{blattmann2023align}, and Skorokhodov1 \textit{et al.}\cite{skorokhodov2024hierarchical} introduce the cascade paradigm into video generation. Indeed, the cascade pipeline has shown effectiveness in high-resolution, the downside is that escalating resolution significantly increases training expenses and computational load, making such models impractical for most researchers and users.

By combining the principles of discrete prior representations with the learned codebook, Vector-Quantized Variational Autoencoder (VQ-VAE) \cite{razavi2019generating} enables high-quality image \cite{hu2022global}, video \cite{yan2021videogpt}, and speech \cite{chen2024stylespeech} generation. Building on this, CodeFormer \cite{zhou2022towards} uses a learned discrete codebook for blind face restoration. 
 FlowVQTalker \cite{tan2024flowvqtalker} develops a Vector-Quantized Image Generator to enhance emotion-aware textures and clear teeth. 
 In this work, we extend VQ-VAE as an image render network to achieve high-resolution image generation in an end-to-end manner.

\section{Method}
In this paper, we propose a two-stage framework for generating high-resolution talking faces only with speech inputs. The overview of our framework is illustrated in Fig. \ref{fig:framework}. This framework comprises speech-to-portrait generation and speech-driven talking face two stages. In the first stage, the speaker's portrait is generated based on the speech-portrait correlation. And then the generated portrait is used as a reference image to synthesize high-resolution talking face videos in the next stage. 
Therefore, we first introduce the speech-to-portrait generation in Section \ref{sec:SCFP}, and talking face generation in Section \ref{sec:HRTF}.

\subsection{Speech-Conditioned Portrait Generation with Face Prior Guidance }
\label{sec:SCFP}


\subsubsection{Observation and Motivation}
Conditional LDMs are powerful generation models capable of synthesizing results aligned with the given condition. Previous study \cite{katospeech} leverages the capability of conditional LDM to generate face portraits from speech input. Specifically, during the training phase, the face image is embedded into latent representation $z^f$ (referred to as $z^0$ in the diffusion process) by a pre-trained face encoder $E_{f}$, while the speech condition is represented as feature $z^s$ extracted by a pre-trained speech encoder $E_{s}$. Then the face embedding is destroyed into a noised vector $z^t$, characterized by a Gaussian distribution, through a series of 
$t$ time steps in the diffusion process, which is denoted as:
\begin{equation}
    z^t:=\alpha^t*z^0+(1-\alpha^t)*\epsilon,
    \label{eq:zt}
\end{equation}
where $\epsilon \sim \mathcal{N}(\textbf{0},\textbf{I})$ denotes the injected noise, and $\alpha^t$ represents the noise level at the $t$ time step. 
 Based on speech-face pairs, the conditional LDM is trained via 
\begin{equation}
    L_{LDM} := \mathbb{E}_{\epsilon,  z^{t},  t} \left[\left\|\epsilon-\epsilon_\theta\left(z^s, z^{t}, t\right)\right\|^2\right],
    \label{eq:LDM}
\end{equation}
where $\epsilon_\theta$ denotes the optimized denoising model, and $\theta$ denotes its parameters.
During the inference process, $z^{t}$ is sampled from Gaussian distribution $\mathcal{N}(\textbf{0},\textbf{I})$, samples from denoised result $\Tilde{z}^0$
are decoded to image space with the pre-trained decoder, which is obtained through
\begin{equation}
    \Tilde{z}^0 := \frac{1}{a^t}(z^t-(1-a^t)*\epsilon_\theta\left(z^s, z^{t}, t\right).
    \label{eq:denoiseZ0}
\end{equation}

\begin{figure}[t]
  \centering
  \includegraphics[width=1\columnwidth]{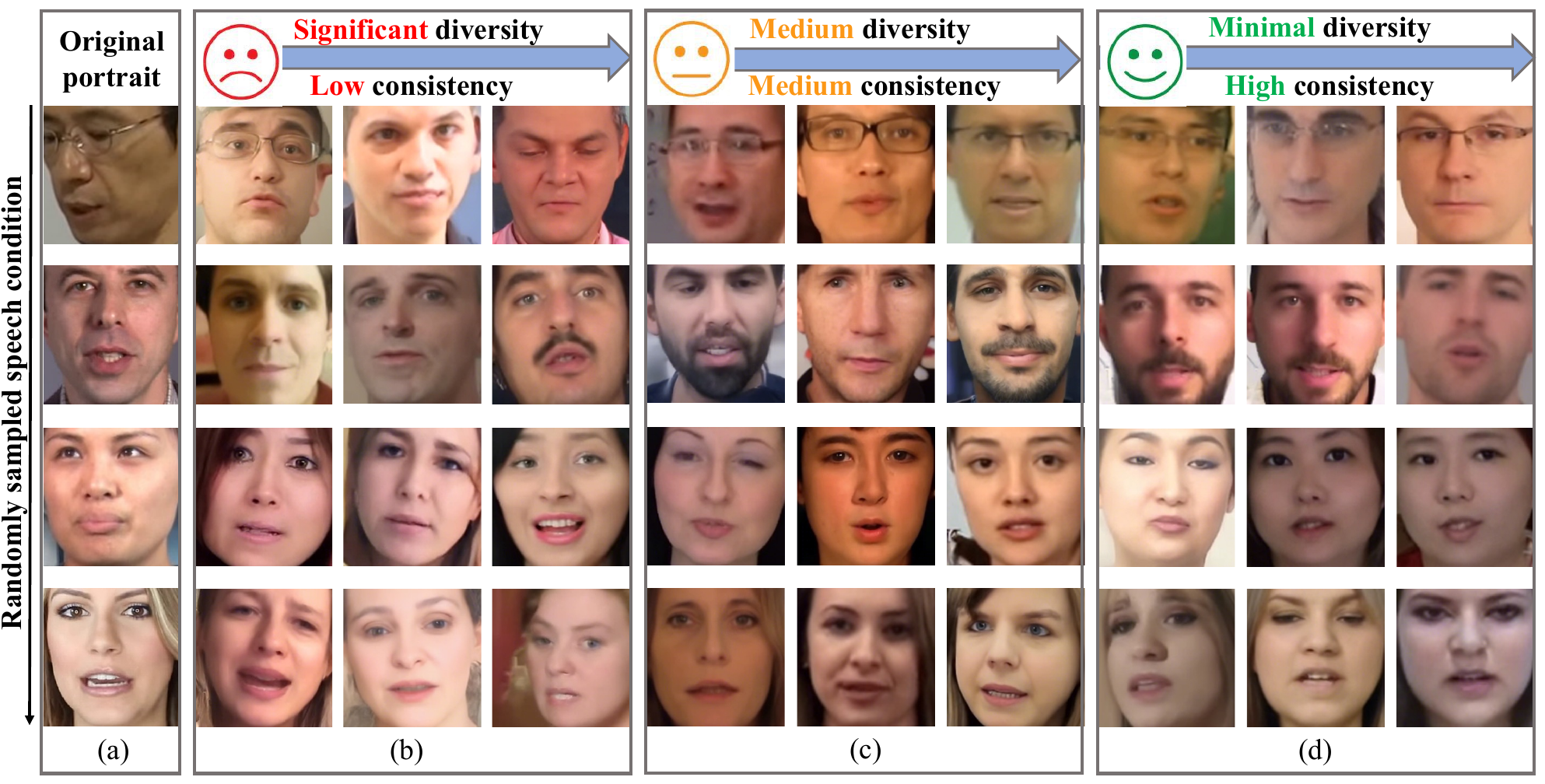}
  \caption{ Qualitative comparison of speech-conditioned portrait generation without or with statistical face prior guidance. (a) Ground truth cropped from the video frame; (b) Top-3 generated results of the same speech condition without face prior guidance; (c) Top-3 generated results of the same speech condition with sample-equivalent weighted ($\beta^0$) face prior guidance; (d) Top-3 generated results of the same speech condition with sample-adaptive weighted ($\beta$) face prior guidance. \textbf{Diversity} refers to the variance among the generated results of different sample noise with the same speech condition, while \textbf{consistency} denotes the preservation of identity in generated results compared to the ground truth.}
  \label{fig:figure_prior}
  \vspace{-5pt}
\end{figure}

\begin{figure}[b]
    \centering
    \includegraphics[width=0.5\linewidth]{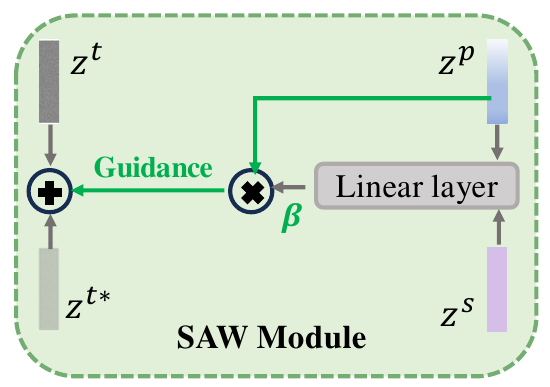}
    \caption{The details of proposed Sample-adaptive weighted module (SAW).}
    \label{fig:SAW}
\end{figure}

\begin{figure*}[htbp]
    \centering
    \includegraphics[width=0.9\linewidth]{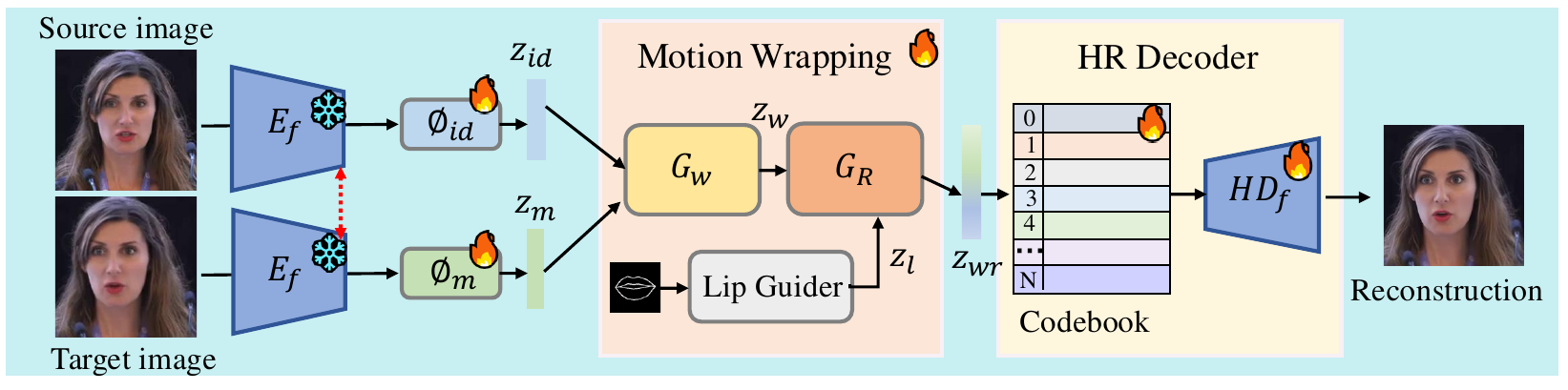}
    \caption{The details of holistic motion construction and wrapping.  
    We train identity encoder $\varnothing_{id}$, motion encoder $\varnothing_{m}$, Motion Wrapper, and HR Decoder to learn holistic motion representation and motion wrapping.   }
    \label{fig:motion}
\end{figure*}

 Although S2P has seen improvements with the adoption of LDM, leveraging the speech condition to precisely generate the corresponding image remains challenging due to the inherent high levels of output diversity in LDM.  This problem is evident in Fig. \ref{fig:figure_prior}, where outputs conditioned on the same speech clip exhibit significant diversity in characteristics.
   We attribute the failure to generate accurate and realistic face portraits to two main factors: \textbf{(i)}
   All desired aspects of a face portrait must be conveyed solely by the input speech signal, which inherently contains limited information, making it difficult to precisely convey all necessary details.
   \textbf{(ii)} The generation process (\textit{i.e.,} denoising process) begins with a randomly sampled Gaussian noise, which does not contain any information about the target face image, thus making it very challenging for the LDM to accurately reconstruct the exact facial features from scratch. 
  Therefore, the generation results are usually with significant diversity and low consistency, as shown in Fig. \ref{fig:figure_prior} (b).

\subsubsection{Conditional LDM with Face Prior as Guidance}

    Based on the fact that the skeletal structure of the human face is generally the same, this naturally results in the statistically average face feature as useful information for S2P. Therefore, we propose a formulation for the portrait feature $z^0$ as a combination of the statistical average face feature (\textit{i.e., }statistical face prior) $z^p$ and and personalized facial variance $z^v$:
 \begin{equation}
    z^0 = z^p + z^v.
     \label{eq:Z0}
 \end{equation}
Instead of starting from a random noise only with speech conditions, 
we propose introducing the statistical face prior into the random noise to provide general structural information. This approach reformulates the generation of the portrait latent code $z^0$ in the denoising process as the generation of facial variance $z^v$ implicitly.

To obtain the statistical face prior, as shown in Fig. \ref{fig:overview}, we statistically average the features extracted from the pre-trained face encoder $E_{f}(\cdot)$ on a given dataset with gender balance:
\begin{equation}
    z^{p} = \frac{1}{N} \sum_{i=1}^{N} E_{f}(f^i),
\end{equation}
where $f^i$ denotes the $i$-th face image and $N$ is the total number of face images. 
Through experiments, we observe that as $N$ gradually increases, the statistical face prior tends to converge, suggesting that the calculated prior becomes representative of the shared characteristics. Here, we set $N=10000$ in this work.

Formally, given the calculated face prior $z^p$, we add it with the noised image latent code $z^t$, yielding the input $z^{t*}$ for denoise UNet.
As illustrated in Fig. \ref{fig:overview}, we modify the noisy representation $z^t = \epsilon \sim \mathcal{N}(\textbf{0},\textbf{I})$ 
   into 
   \begin{equation}
       z^{t*} = z^p + \epsilon \sim \mathcal{N}
   (z^p,\textbf{I}).
   \label{eq:mixture input}
   \end{equation} 
Therefore, the learning objective of portrait diffusion can be defined as:
      \begin{equation}
    L_{P_{diff}} := \mathbb{E}_{\epsilon, z^{t*}, t} \left[\left\|\epsilon-\epsilon_\theta\left(z^s, z^{t*}, t\right)\right\|^2\right].
    \label{eq:LDM with prior}
\end{equation}

By guiding the denoising process with an explicit face prior, 
we provide prompt information about the basic structures, thereby enabling the model to focus more on personalized facial variance. This, in turn, facilitates the alignment between the generated face portrait and the speech condition. 
However, in real-world scenarios, individuals with similar speech characteristics may present different facial attributes.

Inspired by prior works on personalized modulation in speech-conditioned generation~\cite{choi2022snac, yu2024gaussiantalker}, we propose a lightweight Sample-Adaptive Weighting (SAW) module that dynamically modulates the statistical face prior according to the input speech. Although simple in form, this linear design effectively functions as an attention-like mechanism, enhancing identity-related facial features in a sample-dependent manner.

As shown in Fig. \ref{fig:SAW}, given the speech latent code \( z^s \) and the statistical face prior \( z^p \), the SAW module computes modulation weights as follows:
\begin{equation}
    \beta = \text{Linear}([z^s, z^p]) = \mathbf{W}_s z^s + \mathbf{W}_p z^p + \mathbf{b},
    \label{eq:weight}
\end{equation}
where \( \mathbf{W}_s \) and \( \mathbf{W}_p \) are learned projections. This formulation enables the speech signal to inject sample-specific preferences while recalibrating the prior attributes within a shared modulation space. The resulting weights \( \beta \) act as dimension-wise gates, modulating the face prior through element-wise multiplication:
\begin{equation}
    z^{t*} = \beta \odot z^p + \epsilon, \quad \epsilon \sim \mathcal{N}(0, \mathbf{I}),
    \label{eq:weighted_face_prior} 
\end{equation}
producing a speech-adaptive prior distribution that is used in both training and inference within the diffusion model.
This design achieves efficient and content-aware feature selection without requiring explicit softmax normalization or attention mechanism, providing a computationally lightweight yet effective mechanism for speaker identity preservation.

\subsubsection{Contrastive and Reconstruction Pre-training}
Cross-modal alignment representation learning is crucial for S2P since the speech signal drives the synthesis of portrait images. Inspired by the great success of contrastive learning in various cross-modal applications \cite{afham2022crosspoint,parelli2023clip}, we employ contrastive learning in this section to facilitate speech-face alignment. 
However, contrastive learning, while explicitly leveraging useful audiovisual pair information, may discard modality-unique information that is valuable in portrait generation.
Additionally, the reconstruction task may force its representation to encode pixel details. 
 The complementary of these two representation learning paradigms motivates us to integrate them for aligned and detailed speech-portrait representation.
 
 In this section, we propose \textbf{Con}trastive and \textbf{Re}construction (ConRe) pre-training that unifies the two representation learning paradigms.
 Given a speech clip of a speaker and the corresponding face portrait, we employ a speech encoder $E_{s}(\cdot)$ and a face encoder $E_{f}(\cdot)$ to extract the speech embedding $z^s \in \mathcal{R}^d$ and the face embedding $z^f \in \mathcal{R}^d$, respectively, where $d$ is the dimension of the embedding vectors.
To align the speech embedding and face embedding, a symmetric cross-entropy loss \cite{radford2021learning} ($L_c$) is applied, leveraging contrastive learning techniques. Specifically, we use VGGFace \cite{qawaqneh2017deep} as the face encoder, and a model combined with CNN (the speech encoder architecture in Speech2Face \cite{oh2019speech2face}) and Convolutional Block Attention Module (CBAM) \cite{woo2018cbam} as the speech encoder. Since we apply the diffusion model in a latent space, a face decoder is required to upsample the latent representation into image space.
A CNN-based model symmetrical to the VGGFace is designed as the face decoder $D_{f}(\cdot)$. 
Finally, a combination of MAE loss and Learned Perceptual Image Patch Similarity (LPIPS) loss \cite{zhang2018unreasonable} is used as the reconstruction loss ($L_r$). 
The objective function of the CoRe pre-training $L_{CR}$ is defined as:
\begin{equation}
    L_{CR} = L_c + L_r,
\end{equation}
where $L_{c}$ and $L_r$ denote the contrastive loss and the reconstruction loss, respectively.

\subsection{High-Resolution Talking Face Synthesis with Holistic Motion and Lip Region Refinement}
\label{sec:HRTF}

After identifying the speaker, the generated portrait is used to provide identity information in the next talking face generation process. Rather than directly generating video frames, we aim to estimate holistic motion in the latent space conditioned on the speech. To achieve this, we first construct the motion latent space and train the encoder,  decoder, motion learner, and motion wrapping network. Subsequently, we train a motion diffusion model to capture the learned motion distribution conditioned on speech, enabling the generation of motion latent variables during inference.

\subsubsection{Holistic Motion Construction and Wrapping}

Given a corpus of talking face videos, we aim to build a motion latent space for speech dynamics and a wrapping mechanism for video frame generation. 
As shown in Fig. \ref{fig:motion}, 
we first randomly select two frames from the same video, a source image $I_s$ and a target image $I_t$. Image encoder $ E_f$ then encodes $I_s$ and  $I_t$ as latent maps. An identity encoder $\varnothing_{id}$ is used to extract identity information $z_{id}$ from the latent map of the source image, while a motion encoder $\varnothing_{m}$ is used to learn motion code $z_m$ from the latent map of the target image.
These processes can be defined as:
\begin{align}
    z_{id} &= \varnothing_{id}(E_f(I_s)); \\
    z_{m}  &= \varnothing_{m}(E_f(I_t)). \nonumber
\end{align}
Then the extracted identity information $z_{id}$ and motion code $z_m$ are input into the motion wrapping module to transform the learned motion into the identity speaker.
Motion wrapping module comprises a motion wrapper $G_w$ and a lip refiner $G_R$. The motion wrapper $G_w$ is a flow predictor, which takes $z_{id}$ and $z_m$ as input to estimate the latent flow from $I_s$ to $I_t$. However, only using latent flow to warp latent may be insufficient to generate the latent map of $I_t$ due to the occlusions in some positions of $I_s$ \cite{siarohin2019first}, we follow \cite{ni2023conditional,wang2024lia} to also estimates a latent occlusion map $o$ in the flow predictor. Latent occlusion map $o$ contains values changing from 0 to 1 to indicate the degree of occlusion, where 1 is not
occluded and 0 means entirely occluded. The wrapped latent map $z_w$ can be produced by:
\begin{equation}
    z_w = o \odot \tau(z_m,z_{id}),
\end{equation}
where $\odot$ denotes the Hadamard product and $\tau$ denotes warping operation.
To enhance the lip movement in the wrapped latent map $z_w$, we design a lip refiner $G_R$ to explicitly learn the lip guidance $z_l$, which is produced by the lip guider with lip landmark as input. 
The lip landmark is generated by a fine-tuned audio2lmk module in \cite{wei2024aniportrait}. Therefore, the final wrapped latent map $z_{wr}$ can be generated through:
\begin{equation}
    z_{wr} = G_R(z_w, z_l).
\end{equation}
 Decoder $HD_f$ subsequently decodes the final wrapped latent map $z_{wr}$ to reconstruct target image $\hat{I}_t$. 
 Therefore, this pipeline can be trained with the following loss:
 \begin{equation}
     L = L_{re}+ L_{vgg} + L_{adv},
     \label{eq:L_recon}
 \end{equation}
where $L_{re}$, $L_{vgg}$, and $L_{adv}$ denote a reconstruction loss, a perceptual loss, and an adversarial loss, respectively. 
 $L_{re}$ is calculated to minimize the pixel-wise distance, which can de defined as:
 \begin{equation}
     L_{\text{re}}(I_t,\hat{I}_t) = \mathbb{E}\left[\|I_t - \hat{I}_t\|_1\right].
 \end{equation}
 Towards minimizing the perceptual distance, we apply a VGG19-based  $L_{vgg}$ on multi-scale feature maps between ground truth and reconstruction, written as:
 \begin{equation}
     L_{\text{vgg}}(I_t,\hat{I}_t) = \mathbb{E}\left[\sum_{n=1}^{N} \|F_n(I_t) - F_n(\hat{I}_t)\|_1\right],
 \end{equation}
 where $F_n$ denotes the $n_th$ layer in a pre-trained VGG19 \cite{simonyan2014very}. Further, towards generating photo-realistic results, we incorporate an adversarial loss $L_{adv}$, which is calculated as:
 \begin{equation}
     L_{\text{adv}}(\hat{I}_t) = \mathbb{E}_{\hat{I}_t\sim p_{\text{rec}}} \left[ -\log(D(\hat{I}_t)) \right],
 \end{equation}
 where $D$ is a discriminator.
 
High resolution is required for generated video frames, instead of employing the costing two-stage resolution scale-up paradigm, we are inspired by
the advantages of discrete codebook prior in image restoration task \cite{van2017neural,zhou2022towards}, we propose to adopt a codebook to scale up resolution.

Given the sparsity of high-resolution video data, we fine-tune the discrete codebook in the work of Zhou \textit{et al.} \cite{zhou2022towards} and the decoder $HD_f$ to store high-resolution visual parts of face images via self-reconstruction learning. We map the final wrapped latent map $z_{wr}$ with the nearest item in the learnable codebook $C = \left\{ c_k \in \mathbb{R}^d \right\}_{k=0}^{N}$ to obtain the quantized feature $z_q \in \mathbb{R}^{m \times n \times d}$ via:
\begin{equation}
    z_q = Q(z_{wr}) = \arg \min_{c_k \in C}\|z_{wr}^{(i,j)} - c_k\|_2^2,
\end{equation}
where $z_{wr}^{(i,j)}$ denote the $(i,j)$ ``pixel" of $z_{wr}$.
Then, we adopt the intermediate code-level loss $L_{code}$ and image-level reconstruction losses denoted in Eq. \ref{eq:L_recon}, to supervise self-reconstruction. $L_{code}$ is calculated to to reduce the distance between codebook $C$ and the final wrapped latent map $z_{wr}$:
\begin{equation}
   L_{code} = \| \text{sg}(z_{wr}) - z_q \|_2^2 + \beta \| z_{wr} - \text{sg}(z_q) \|_2^2.
\end{equation}

By learning from frame-to-frame reconstruction, we can obtain a holistic motion construction and high-resolution wrapping modules.

\subsubsection{Holistic Motion Generation with Diffusion Model}

Given the constructed holistic latent space, we can extract the facial dynamics and head movements from real-life talking face videos, to train a motion diffusion $M_{diff}$ to learn the distribution of latent motion with the speech condition. 

As illustrated in Fig. \ref{fig:overview}, a latent motion sequence extracted from a video clip is defined as $\textbf{M} = \{z^i_m, \dots, z^N_m\}$, where $N$ is the number of video frames.
During the training process of $M_{diff}$, $\textbf{M}$ is gradually converted to Gaussian noise $\textbf{M}_t$, where $t$ denotes the number of total denoising steps. Additionally, the accompanying speech clip $S$ is fed into a pretrained feature extractor to extract features $z^s$. And then $M_{diff}$ is trained to eliminate noise from the Gaussian noise condition on speech features:
\begin{equation}
  M_{diff} = \mathbb{E}_{t, M, \epsilon} \left[ \| \epsilon - \hat{\epsilon}_t(M_t, t, z^s) \|_2^2 \right] .
\end{equation}
This iterative process better captures the distribution of motion.

\subsection{Inference}
At inference time, given an arbitrary speech clip, \( P_{\text{diff}} \) predicts the face portrait using the speech-face correlation to identify the speaker in Stage 1. The generated face portrait is then edited by Deep Live Cam \footnote{https://github.com/hacksider/Deep-Live-Cam}, which alters nonrelated speech attributes while preserving facial consistency. This edited portrait serves as the reference image for synthesizing the talking face video to visualize the speech dynamics.
In Stage 2, \( M_{\text{diff}} \) estimates the holistic motion, while the fine-tuned audio2lmk model generates the lip landmarks from the source speech. The motion-wrapping module then adapts the estimated motion to the target speaker in the edited portrait, refining the lip movement. Finally, video frames are generated using our high-resolution decoder.

 \begin{table*}[htbp]
 \centering
     \caption{Comparison results on AVSeech dataset. The best results are highlighted in bold. Note that $\downarrow$ indicates that a smaller value is preferable, while $\uparrow$ indicates that a larger value is preferable.}
    \begin{tabular}{l|c|ccc|cc|ccc}
    \hline
    
             \multirow{2}{*}{Method} &\multirow{2}{*}{Year}& \multicolumn{3}{c}{Feature Similarity} \vline & \multicolumn{2}{c}{Identity Preservation} \vline & \multicolumn{3}{c}{Retrieval Performance} \\
    \cline{3-10}
         & & L1 $\downarrow$ & L2 $\downarrow$ & cos $\downarrow$ &
          gender $\left (\%\right) \uparrow$ & age $\left (\%\right) \uparrow$ &$R@1 \uparrow$ & $R@2 \uparrow$ &$R@5 \uparrow$\\
    \hline
    Wav2Pix \cite{duarte2019wav2pix}&2019&144.72& 24.32&82.51& 67.4&41.3 &2.46&6.72& 14.26 \\
    Speech2Face \cite{oh2019speech2face}& 2019&67.18&3.94 &46.97 &95.6 &65.2 &9.17&14.94&28.31 \\
    Choi \textit{et al.} \cite{choi2019inference}&2019& 60.26&3.57 &35.89 &95.8 &69.6 &10.84 &17.37 & 32.91\\
    SF2F \cite{bai2022speech}& 2022&89.31 &17.49 &64.83 &72.1 & 48.9&7.37 &13.45 &20.72 \\
    Kato \textit{et al.} \cite{katospeech}& 2023&46.35& 2.73&21.96 & 96.7&81.3 &18.44 & 28.31&49.24\\
    SCLDM (Ours)&-& \textbf{31.26}&\textbf{1.14}&\textbf{10.35} & \textbf{99.1} &\textbf{86.4} & \textbf{21.45}& \textbf{36.21}& \textbf{59.86}\\
    \hline
    \end{tabular}

    \label{tab:comparison resuls on AVSpeech}
\end{table*}

 \begin{table*}[htbp]  
    \centering
    \caption{Comparison results on VoxCeleb dataset. The best results are highlighted in bold.  }
   
    \begin{tabular}{l|c|ccc|cc|ccc}
    \hline
    
        \multirow{2}{*}{Method} &\multirow{2}{*}{Year}& \multicolumn{3}{c}{Feature Similarity} \vline & \multicolumn{2}{c}{Identity Preservation} \vline  & \multicolumn{3}{c}{Retrieval Performance}\\
    \cline{3-10}
         & & L1 $\downarrow$ & L2 $\downarrow$ & cos $\downarrow$ &
          gender $\left (\%\right) \uparrow$ & age $\left (\%\right) \uparrow$&$R@1 \uparrow$ & $R@2 \uparrow$ &$R@5 \uparrow$ \\
    \hline
    Wav2Pix \cite{duarte2019wav2pix}&2019&137.58 &22.19 &79.36 & 74.5&49.6 &4.81& 9.56&12.94 \\
    Speech2Face \cite{oh2019speech2face}&2019& 66.46&2.77 &44.38 &96.1 & 69.4&7.79& 14.38&20.14 \\
    Wen \textit{et al.} \cite{wen2019face}&2019& 59.82& 2.41& 42.54&97.4 & 72.5 &8.26& 15.62& 23.51\\
    Choi \textit{et al.} \cite{choi2019inference}&2019& 56.32&2.24&30.49 &97.6 &74.8&9.43 &16.32 & 28.67\\
    SF2F \cite{bai2022speech}&2022&78.45 & 13.31&58.79 &79.3 & 57.6&9.25 &17.17 &22.53 \\
    Kato \textit{et al.} \cite{katospeech}&2023& 40.11& 2.26&18.74 & 98.1&83.8 &16.19 & 25.64&42.38\\
    SCLDM (Ours)& -&\textbf{25.24} & \textbf{0.91} &\textbf{9.86} &\textbf{99.6} &\textbf{89.3} & \textbf{19.84}& \textbf{33.57}& \textbf{51.79} \\
    \hline
    \end{tabular} 
    \label{tab:comparison results on VoxCeleb}
\end{table*}

 \begin{figure}[htbp]
\centering
\includegraphics[width=0.98\columnwidth]{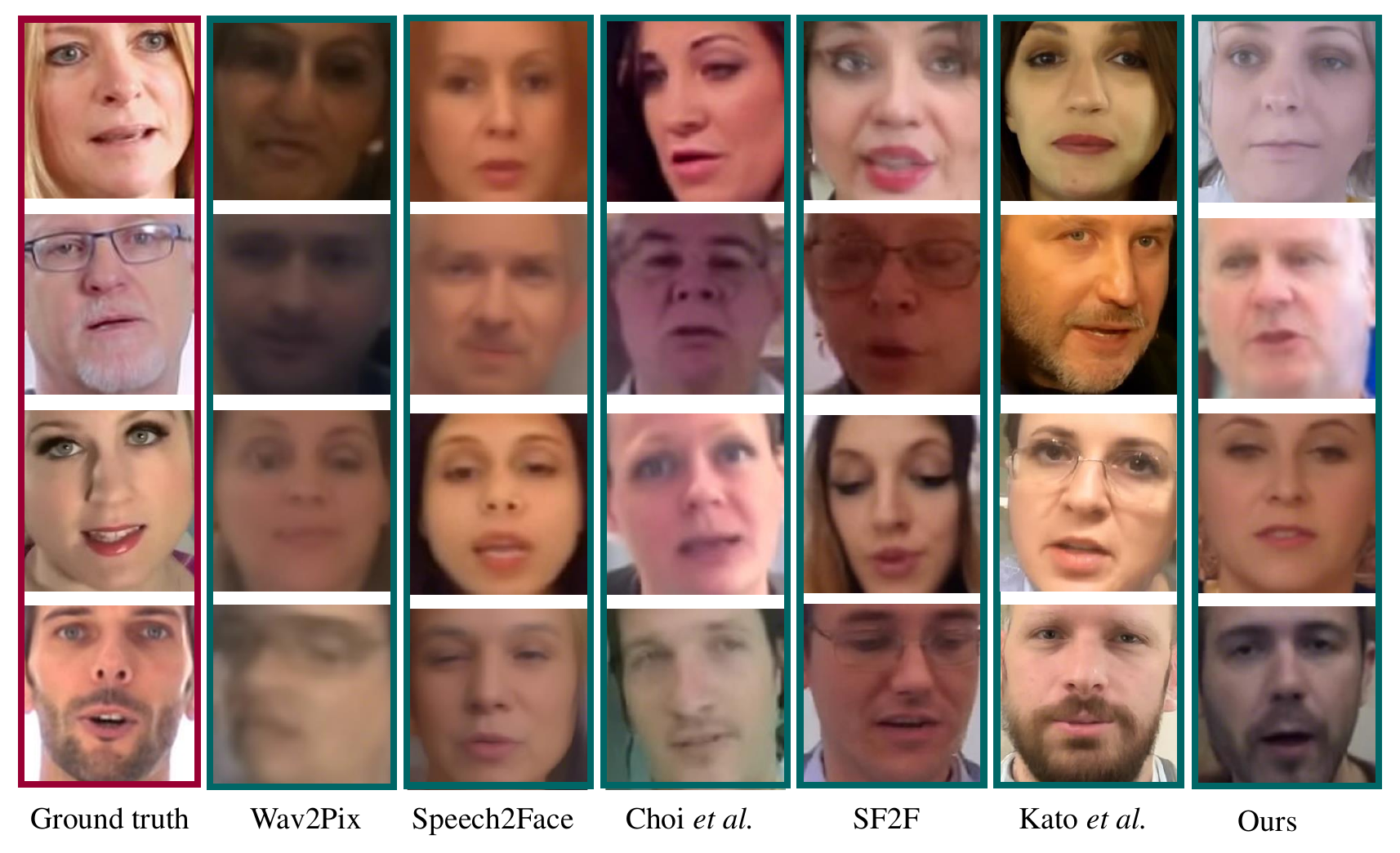} 
\caption{Qualitative comparison between our model and previous SOTA methods on the AVSpeech dataset. }
\label{fig:comparative_Avspeech}
\vspace{-8pt}
\end{figure}

\section{Implementations}

\subsection{Datasets}
For \textbf{speech-to-portrait}, we empirically validate the effectiveness of our proposed method on the AVSpeech \cite{ephrat2018looking} and VoxCeleb \cite{nagrani2017voxceleb} datasets. The AVSpeech dataset is a large-scale audio-visual collection from YouTube, comprising 2.8 million video clips. It features significant diversity, including face images extracted from videos captured "in the wild." The VoxCeleb dataset contains 1,251 speakers spanning a wide range of ethnicities, accents, professions, and ages, with additional metadata on each speaker’s nationality and gender.

For \textbf{talking face generation}, we leverage three widely-used datasets: VoxCeleb \cite{nagrani2017voxceleb} and HDTF \cite{zhang2021flow}. HDTF is a large, in-the-wild, high-resolution, and high-quality audio-visual dataset comprising approximately 362 videos, totaling 15.8 hours. The original video resolutions are 720P or 1080P. 
Additionally, to further assess the generalizability of our talking face generation approach, we collect 50 English videos from the AVSpeech \cite{ephrat2018looking} test set as a wild dataset for evaluation.

\subsection{Data Preprocessing}
 \textbf{Image processing:} We utilized Dlib \cite{king2009dlib}, a publicly available software, to detect the face in the first frame of the video clips, if more than one face were detected, a face closer to the  coordinates of the target speaker was selected. Note that the coordinates of the target speakers are provided in the AVSpeech dataset. The images we cropped were all resized to 256 $\times$ 256. These procedures of cropping and resizing were adapted to all images in the VoxCeleb dataset.

 \begin{figure}[htbp]
\centering
\includegraphics[width=1\columnwidth]{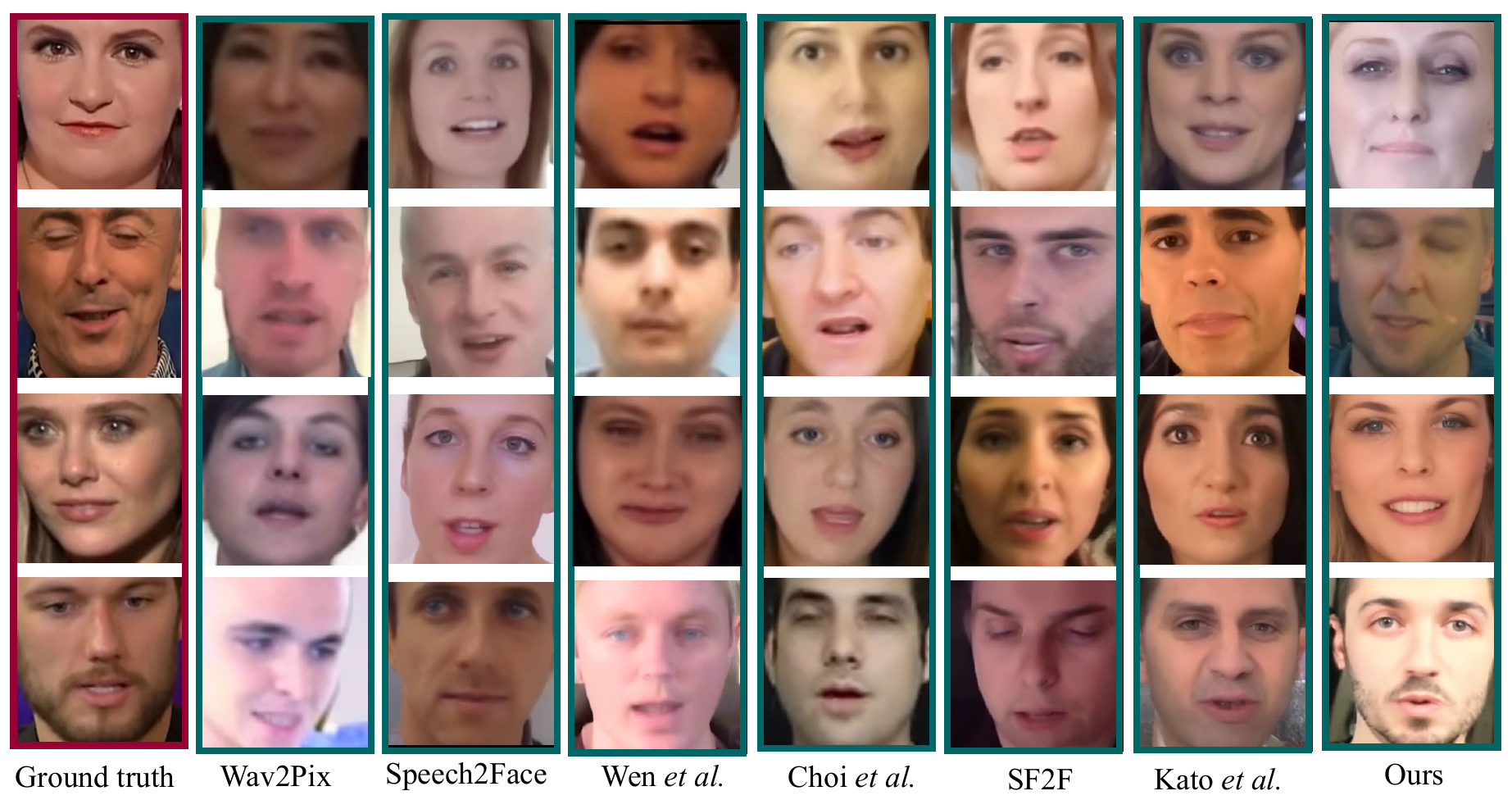} 
\caption{Qualitative comparison between our model and previous  SOTA methods on the Voxceleb dataset. }
\label{fig:comparative_voxceleb}
\end{figure}

 \noindent\textbf{Audio processing:} All audio samples were separated from the corresponding video clips and then resampled to 16kHz. Following the previous works \cite{choi2019inference,oh2019speech2face}, we used 6 seconds of audio, if the audio was longer than 6 seconds, it was truncated, while if the audio was shorter than 6 seconds, it was duplicated until it became longer than 6 seconds. And then it was truncated to be 6 seconds. We calculated the spectrograms of the audio sample by taking STFT with a Hann window of 25 mm, a hop length of 10 ms, and 512 FFT frequency bands. Each complex spectrogram $S$ subsequently went through the power-law
compression, resulting $sgn(S)|S|^{ 0.3 }$ for real and imaginary independently, where $sgn(.)$ denoted the signum.
 \noindent\textbf{Video processing:} 
 All video frames of three datasets, whether for training or testing, are resized into $256 \times 256$. Additionally, we excluded faces with resolutions lower than $256 \times 256$. The videos are sampled at 25 FPS, and the audio is pre-processed to 16 KHZ.

 \begin{table*}[htbp]
    \centering
    \caption{Ablation results on AVSpeech dataset. The best results are highlighted in bold. }
   
    \begin{tabular}{ccc|ccc|cc|ccc}
    \hline
    
        \multicolumn{3}{c}{Method} \vline& \multicolumn{3}{c}{Feature Similarity} \vline & \multicolumn{2}{c}{Identity Preservation} \vline& \multicolumn{3}{c}{Retrieval Performance}  \\
    \hline
         Baseline  &  ConRe & SAW & L1 $\downarrow$ & L2 $\downarrow$ & cos $\downarrow$ & gender $\left (\%\right) \uparrow$ & age $\left (\%\right) \uparrow$&$R@1 \uparrow$ & $R@2 \uparrow$ &$R@5 \uparrow$ \\
    \hline
     $\checkmark$& & &56.39 &3.30&29.83&95.9 & 74.7&12.82& 19.65& 34.81\\
    $\checkmark$&$\checkmark$& &44.27 &2.38 &20.41&96.4 & 80.3&18.97&29.32&49.96 \\
    $\checkmark$ &  & $\checkmark$& 42.14& 2.19&18.74 &96.8&81.5&15.21&24.63&42.77 \\
    $\checkmark$&$\checkmark$ &$\checkmark$& \textbf{31.26}& \textbf{1.14} &\textbf{10.35} &\textbf{99.1} &\textbf{86.4}& \textbf{21.45}& \textbf{36.21}& \textbf{59.86}  \\
    \hline
    \end{tabular}
    \label{tab:ablation results on VoxCeleb}
\end{table*}


\begin{figure}[htbp]
\centering
\includegraphics[width=0.8\columnwidth]{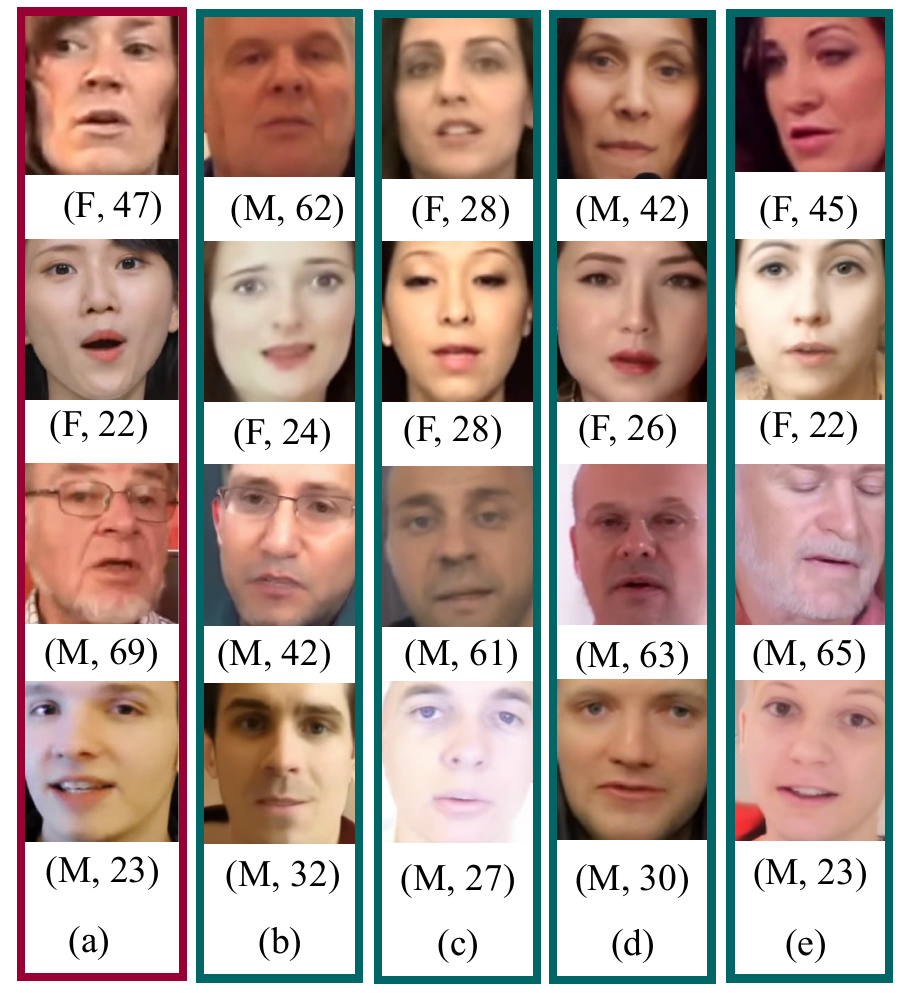} 
\caption{Qualitative comparison of ablation studies with attributes (gender, age), "F" and "M" mean female and male, respectively. (a) Ground truth cropped from the video frame. (b) Generated images by speech-conditioned LDM (Baseline). (c) Generated images by speech-conditioned LDM with ConRe. (d) Generated images by speech-conditioned LDM with SAW. (e) Generated images by SCFP (Ours).}
\label{fig:ablation_cl}
\vspace{-8pt}
\end{figure}

\subsection{Evaluation Metrics}

For \textbf{speech-to-portrait}, we evaluate generation performance based on feature similarity, identity preservation, and retrieval performance. 
\noindent\textbf{Feature Similarity:} Following \cite{oh2019speech2face}, we measure the cosine, L1 and L2 distances between features extracted from the real face image and the generated face image using VGGFace \cite{qawaqneh2017deep}, a pretrained face recognition network.
\noindent\textbf{Identity Preservation:} We utilize Face++\footnote{\url{https://www.faceplusplus.com/attributes}}, a commercial API for facial attribute recognition, to evaluate attributes such as age and gender. Age classification is considered accurate if the age difference between the generated face image and the ground truth is within 10 years.
\noindent\textbf{Retrieval Performance:} Image retrieval involves analyzing visual content in a large image database to identify images that match the query in terms of semantics or similarity \cite{rehman2012content}. To assess identity preservation, we perform image retrieval using the generated portrait as the query image, calculating cosine distances between features of the generated faces and those in the data set. Retrieval performance is reported using the Recall@K metric, such as \( R@1 \), \( R@2 \), and \( R@5 \), which indicate whether the top \( K \) retrieved images contain a true match \cite{wu2022retrievalguard}.

\begin{figure}[htbp]
    \centering
    \includegraphics[width=0.9\columnwidth]{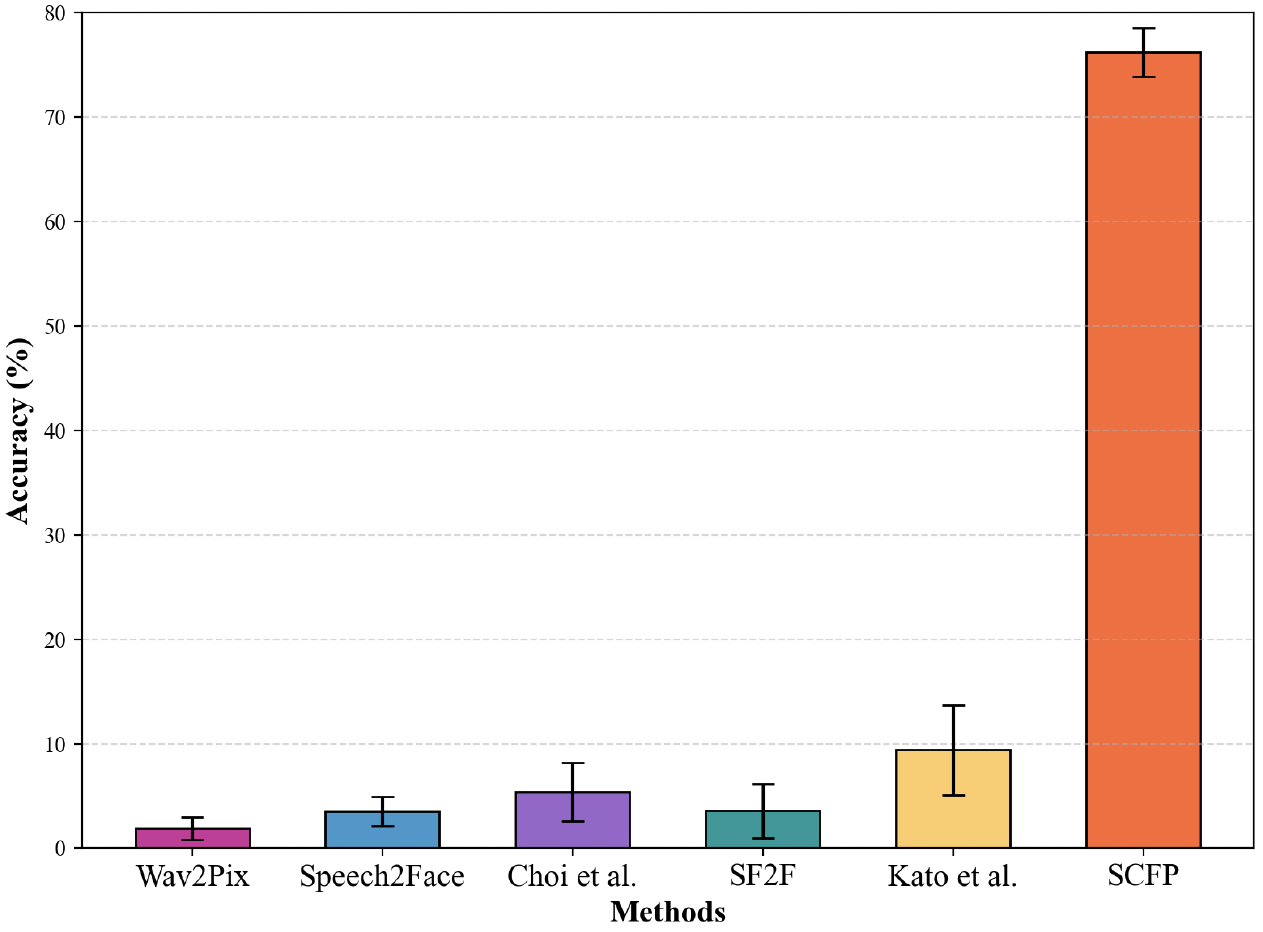}
    \caption{Results of the user study. Among the six methods, our method (SCFP) achieves the highest accuracy for the user evaluation, in terms of image quality and identity preservation.}
    \label{fig:user_study}
\end{figure}

For \textbf{talking head generation}, we evaluate performance based on lip synchronization accuracy and visual quality.
\noindent\textbf{Lip Synchronization Accuracy:} To measure synchronization between speech and lip movements, we use the pre-trained SyncNet model \cite{chung2017out}. The evaluation includes two metrics: Lip Sync Error (LSE-D and LSE-C) \cite{chen2019hierarchical}, where LSE-D calculates the distance between audio and visual features, and LSE-C measures the confidence scores for synchronization.
\noindent\textbf{Visual Quality:} The visual quality of the generated talking head videos is assessed using the Learned Perceptual Image Patch Similarity (LPIPS) metric \cite{zhang2018unreasonable}, which quantifies perceptual differences between generated and ground truth frames. We also compute the Fréchet Inception Distance (FID) \cite{heusel2017gans}, a widely used metric that compares the feature distributions from the Inception network \cite{szegedy2015going} between real and generated images. Additionally, we calculate Structural Similarity (SSIM) \cite{wang2004image} and Peak Signal-to-Noise Ratio (PSNR) to provide a comprehensive evaluation of visual fidelity.
\noindent \textbf{Temporal Consistency:} To evaluate the temporal consistency of generated talking-head videos, we use RAFT \cite{teed2020raft} to extract dense optical flow between consecutive frames and compute the Mean Absolute Difference (MAD) of normalized pixel intensities (range [0,1]) warped along the flow.

 \begin{table*}[htbp]
 \centering
 \caption{Ablation results on face prior weight $\beta$ denoted in Eqn. \ref{eq:weighted_face_prior}. The best results are highlighted in bold.  }
    \begin{tabular}{c|ccc|cc|ccc}
    \hline
    
        \multirow{2}{*}{Face prior weight} & \multicolumn{3}{c}{Feature Similarity} \vline & \multicolumn{2}{c}{Identity Preservation} \vline  &\multicolumn{3}{c}{Retrieval Performance} \\
    \cline{2-9}
          & L1 $\downarrow$ & L2 $\downarrow$ & cos $\downarrow$ &
          gender $\left (\%\right) \uparrow$ & age $\left (\%\right) \uparrow$&$R@1$ & $R@2$ &$R@5$ \\
    \hline
    Sample-equivalent $\beta^0$&35.01&1.48&12.81 &98.8 &84.5 &19.86& 31.41& 51.48 \\
    Sample-adaptive $\beta$& \textbf{31.26}&\textbf{1.14} &\textbf{10.35} &\textbf{99.1} & \textbf{86.4}& \textbf{21.45}& \textbf{36.21}& \textbf{59.86}  \\
    \hline
    \end{tabular}

    \label{tab:Ablation results on prior normalization weight.}
\end{table*}

\begin{table*}[htbp]
    \centering
    \caption{Ablation results on the structural design of face prior weighting mechanism. The best results are highlighted in \textbf{bold}.}
    \label{tab:ablation_structural_design}
    \begin{tabular}{c|ccc|cc|ccc|c}
    \hline
    \multirow{2}{*}{Structural Design} & \multicolumn{3}{c}{Feature Similarity} \vline & \multicolumn{2}{c}{Identity Preservation} \vline & \multicolumn{3}{c}{Retrieval Performance} \vline & \multirow{2}{*}{FLOPs (G)} \\
    \cline{2-9}
    & L1 $\downarrow$ & L2 $\downarrow$ & Cos $\downarrow$ & Gender (\%) $\uparrow$ & Age (\%) $\uparrow$ & R@1 $\uparrow$ & R@2 $\uparrow$ & R@5 $\uparrow$ & \\
    \hline
    Linear    & 31.26 & 1.14 & 10.35 & 99.1 & 86.4 & 21.45 & 36.21 & 59.86 & \textbf{0.0082} \\
    Attention & \textbf{31.20} & \textbf{1.09} & \textbf{10.31} & \textbf{99.1} & \textbf{86.6} & \textbf{21.49} & \textbf{36.27} & \textbf{59.97} & 0.0328 \\
    \hline
    \end{tabular}
\end{table*}

\begin{figure}
    \centering
    \includegraphics[width=0.95\linewidth]{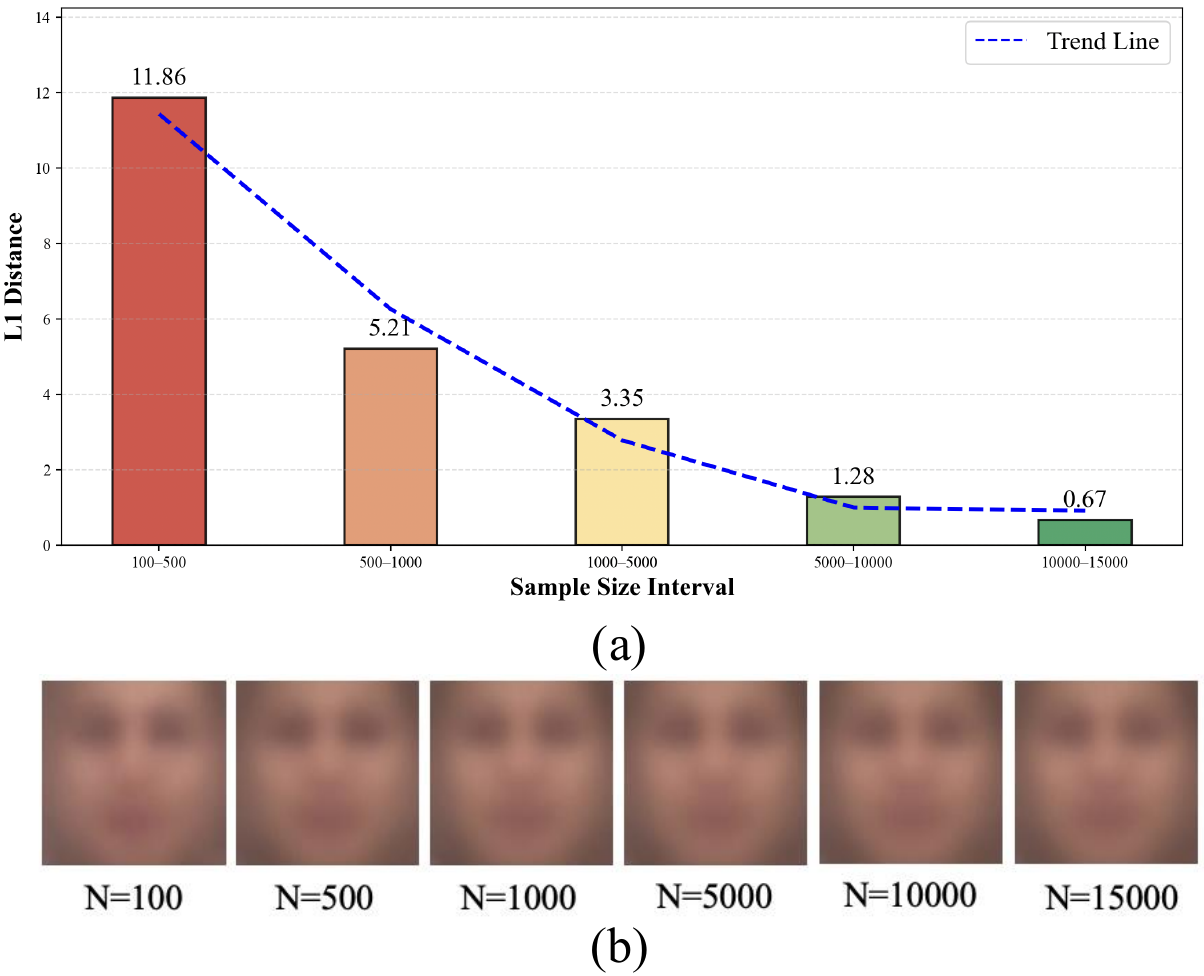}
    \caption{Analysis of the impact of sample size on the statistical face prior. (a) Quantitative comparison. (b) Visualization analysis. }
    \label{fig:face_prior_convegence}
\end{figure}
\subsection{Implementation Details}
The proposed speech-to-portrait generation framework, SCFP, consists of a speech encoder, a face encoder, a face decoder, and a latent diffusion model (LDM). Following Speech2Face \cite{oh2019speech2face}, we use a CNN-based network as speech encoder. To enhance the representation capability of speech features, we incorporate a CBAM module \cite{woo2018cbam} to capture global context. For the face encoder, we employ VGGFace \cite{parkhi2015deep}, while the face decoder is designed as a CNN-based model symmetric to the VGGFace architecture. We adopt the UNet backbone from Stable Diffusion \cite{rombach2022high} as the foundational architecture for LDM in SCFP. 

The proposed talking face generation pipeline, HRTF, comprises a speech encoder, an image encoder, an image decoder, an identity encoder, a motion encoder, a diffusion UNet, a flow generator, a lip refiner, a lip guider, and an audio2lmk module. A pre-trained HuBERT-large model \cite{hsu2021hubert} is used as the speech encoder. The image encoder and decoder are similar to those in LIA \cite{wang2024lia}. Both the identity and motion encoders are implemented using MLPs. The architecture of diffusion UNet is the same as \cite{dhariwal2021diffusion}.
We adopt style blocks from StyleGAN2 \cite{karras2020analyzing} to construct the flow generator. The audio2lmk module and lip guider are adapted from Aniportrait \cite{wei2024aniportrait}.
\vspace{-4mm}

\begin{figure}
    \centering
    \includegraphics[width=0.9\linewidth]{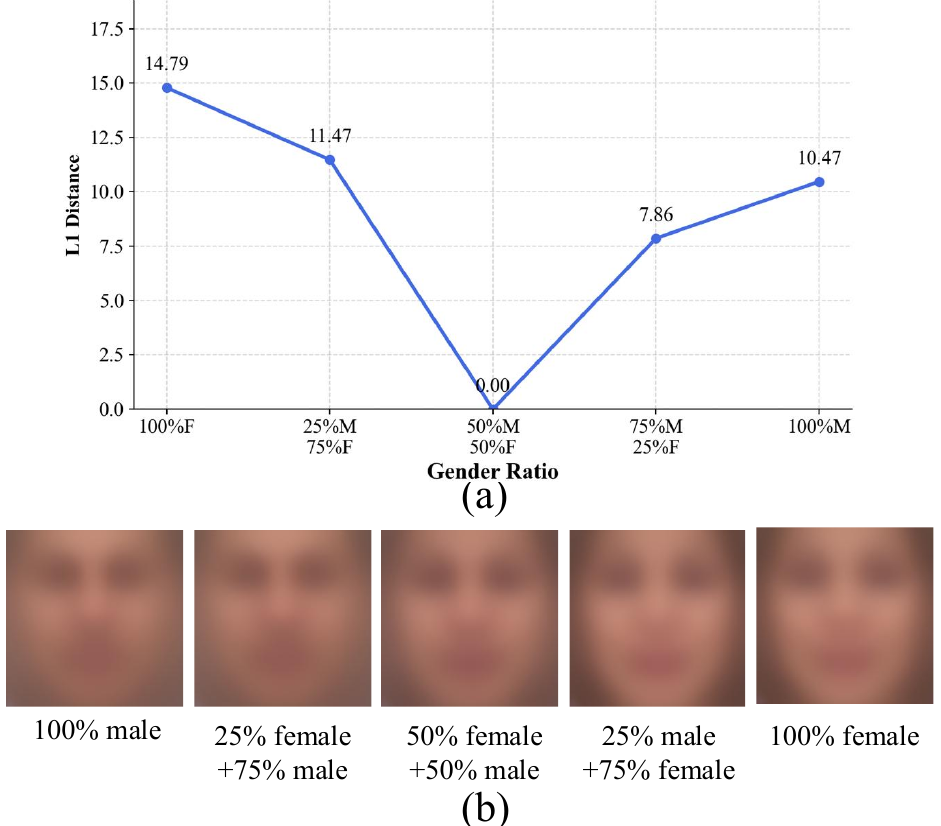}
    \caption{Analysis of the impact of gender ratio on the statistical face prior. (a) Quantitative analysis. (b) Visualization analysis. }
    \label{fig:face_prior_gender}
\end{figure}

\subsection{Training Details}
Our framework is implemented using PyTorch. All experiments were conducted on a GPU server equipped with 8 NVIDIA RTX A6000 GPUs.
For the ConRe pre-training stage, we set the learning rate to 0.0001 for the face encoder and face decoder, and 0.001 for the speech encoder. In the speech-conditioned LDM training stage, the face encoder, face decoder, and speech encoder are frozen, while optimization is performed with a learning rate of 2e-5.
During the motion construction and wrapping training, we use a learning rate of 0.002 and train the model with the Adam optimizer. For the motion diffusion training, we use Adam with a learning rate of 2e-4.
For the high-resolution component, we initialize weights from CodeFormer \cite{zhou2022towards} and use a learning rate of 1e-4, training on the VFHQ dataset as the high-resolution training data.

 \begin{table*}[htbp]
\centering
\caption{Ablation study on the impact of different speech encoders in speech-to-portrait generation. The best results are highlighted in bold.}
\begin{tabular}{l|ccc|cc|ccc}
\hline
\multirow{2}{*}{Method} & \multicolumn{3}{c|}{Feature Similarity} & \multicolumn{2}{c|}{Identity Preservation} & \multicolumn{3}{c}{Retrieval Performance} \\
\cline{2-9}
 & L1 $\downarrow$   & L2 $\downarrow$   & cos $\downarrow$ & gender (\%) $\uparrow$ & age (\%) $\uparrow$ & $R@1$ $\uparrow$   & $R@2$ $\uparrow$  & $R@5$  $\uparrow$ \\
\hline
Wav2Vec  & \textbf{28.24} & \textbf{1.07}  & \textbf{9.94}   & \textbf{99.2}        & \textbf{86.8}     & \textbf{22.02} & \textbf{37.13} & \textbf{60.29} \\
Ours (CNN-based)       & 31.26 & 1.14  & 10.35  & 99.1        & 86.4     & 21.45 & 36.21 & 59.86 \\
\hline
\end{tabular}
\label{tab:speech_encoder_S2P}
\end{table*}

\section{Experiments}
\subsection{Speech-to-Portrait Generation}

\subsubsection{Comparison with SOTA methods.}
 We compare our proposed method with six SOTA S2P methods, categorized into three groups: 1) \textbf{CNN-based} methods, such as Speech2Face \cite{oh2019speech2face} and SF2F \cite{bai2022speech}; 2) \textbf{GAN-based} methods, such as Wav2Pix \cite{duarte2019wav2pix}, Wen \textit{et al.} \cite{wen2019face}, and Choi \textit{et al.} \cite{choi2019inference}; 3) \textbf{LDM-based} method, Kato  \textit{et al.} \cite{katospeech}.
 We perform experiments using the default settings and official implementations for Wav2Pix \cite{duarte2019wav2pix}, Wen \textit{et al.} \cite{wen2019face}, Choi \textit{et al.} \cite{choi2019inference}, SF2F \cite{bai2022speech}, and Kato  \textit{et al.} \cite{katospeech}.
 However, as the code for Speech2Face \cite{oh2019speech2face}, Choi \textit{et al.} \cite{choi2019inference}, and Kato  \textit{et al.} \cite{katospeech} is not available, we reproduce them based on the descriptions provided in their papers.
 Additionally, we only compare with Wen \textit{et al.} on the Voxceleb dataset, as the identity information of speakers is lacking in the AVSpeech dataset.

\begin{table*}[htbp]
\centering
\caption{: Quantitative comparisons with previous speech-driven talking face generation methods on the VoxCeleb dataset. The best results are highlighted in bold.  }
\begin{tabular}{l|c|cc|cccc|c}
\hline
\multirow{2}{*}{Method} &\multirow{2}{*}{Venue} & \multicolumn{2}{c|}{Lip Consistency} & \multicolumn{4}{c|}{Visual Quality} & Temporal Consistency\\
\cline{3-9}
   & & LSE-D $\downarrow$& LSE-C $\uparrow$ & LPIPS $\downarrow$ & SSIM $\uparrow$& PSNR $\uparrow$ & FID $\downarrow$& MAD $\downarrow$ \\
\hline
Aniportait \cite{wei2024aniportrait} &ECCV 2024&9.63 &6.39 &0.13 &0.56 &29.54 & 44.13& 0.0053\\

Real3D-portrait \cite{ye2024real3d} & ICLR 2024&16.60 &3.24 &0.22& 0.54&29.50 & 42.14&0.0062\\

SyncTalk \cite{peng2024synctalk} &CVPR 2024  &11.46 &4.83&0.22 & 0.56&29.41&46.70&0.0064 \\
\hline
Hallo \cite{xu2024hallo} &arXiv 2024.7 & 12.43&5.14 &0.24&0.59 &29.42 &44.82&0.0059 \\

Anitalker \cite{liu2024anitalker} &ACM MM 2024&14.76 &4.58 &0.23&0.59& 28.61 & 43.45&0.0053\\
\hline
VideoRetalking \cite{cheng2022videoretalking} &SIGGRAPH 2022&9.44 &3.34 &0.12&0.62& 29.47 & 39.47& 0.0061\\
\hline
HRTF (Ours)&- & \textbf{6.61}&\textbf{8.65} &\textbf{0.11} &\textbf{0.67}& \textbf{29.66} &\textbf{29.28}& \textbf{0.0042} \\
HRTF-SCFP (Ours)&- & 7.08&8.23 &0.12&0.65 & 29.42&  32.66& 0.0046\\
\hline
\end{tabular}
\label{tab:Vox-tf}
\end{table*}

\begin{table*}[htbp]
\centering
\caption{: Quantitative comparisons with previous speech-driven talking face generation methods on HDTF dataset. The best results are highlighted in bold.  }
\begin{tabular}{l|c|cc|cccc|c}
\hline
\multirow{2}{*}{Method} &\multirow{2}{*}{Venue} & \multicolumn{2}{c|}{Lip Consistency} & \multicolumn{4}{c|}{Visual Quality} & Temporal Consistency\\
\cline{3-9}
   & & LSE-D $\downarrow$& LSE-C $\uparrow$ & LPIPS $\downarrow$ & SSIM $\uparrow$& PSNR $\uparrow$ & FID $\downarrow$ & MAD $\downarrow$ \\
\hline
Aniportait \cite{wei2024aniportrait} &ECCV 2024&8.04 &6.46 &0.11 &0.75 &29.72 & 41.61& 0.0072\\

Real3D-portrait \cite{ye2024real3d} & ICLR 2024&15.78 &3.87 &0.21& 0.74&29.68 & 39.83& 0.0092\\

SyncTalk \cite{peng2024synctalk} &CVPR 2024  &11.07 &5.51&0.12 & 0.73&29.97&42.28 & 0.0078\\

\hline

Hallo \cite{xu2024hallo} &arXiv 2024.7 & 11.36&5.61 &0.14&0.71 &29.52 & 41.32& 0.0099\\

Anitalker \cite{liu2024anitalker} &ACM MM 2024&12.74 &5.06 &0.20&0.69& 29.62& 40.02&0.0057\\
\hline
VideoRetalking \cite{cheng2022videoretalking} &SIGGRAPH 2022&11.53 &4.59 &0.08&0.76& 29.72 & 36.28& 0.0064\\
\hline
HRTF (Ours)&- & \textbf{5.41}&\textbf{9.74} &\textbf{0.06} &\textbf{0.81}& \textbf{30.63} &\textbf{26.34}& \textbf{0.0048} \\
HRTF-SCFP (Ours)&- & 6.86&8.83&0.09&0.78& 30.61&  29.36 & \textbf{0.0051}\\
\hline
\end{tabular}
\label{tab:hdtf-tf}
\end{table*}

 \noindent\textbf{Quantitative Comparison.} The comparison results on AVSpeech and VoxCeleb datasets are reported in Table \ref{tab:comparison resuls on AVSpeech} and Table \ref{tab:comparison results on VoxCeleb}, respectively. Our method outperforms all the competitors in all metrics. Specifically, the cosine distance of our method achieves 10.35 on the AVSpeech test set and 9.86 on the VoxCeleb test set. The gender recognition accuracy achieves 99.1 and 99.6 on the two datasets. These results verify the effectiveness of our approach in producing identity-preserving portraits.
 
\noindent\textbf{Qualitative Comparison.} The qualitative comparison illustrated in Fig. \ref{fig:comparative_Avspeech} and Fig. \ref{fig:comparative_voxceleb} highlights the effectiveness of our approach in generating realistic outputs that align well with the speaker's attributes. This success can be attributed to the integration of face prior guidance and ConRe pre-training into our framework. By leveraging these components, our model demonstrates superior performance compared to previous methods, producing synthesized portraits that closely resemble the speakers.

\noindent\textbf{User Study.}
We conducted a user study involving 40 human evaluators to assess the perceptual effectiveness of the S2P methods. For this study, we randomly selected 60 speech clips from the AVSpeech test set and synthesized portrait images corresponding to each speaker's speech. The evaluators were presented with both the true face and the generated face images and asked to choose the best image based on two criteria: 1) image quality, and 2) identity preservation. As depicted in Fig. \ref{fig:user_study}, the mean and standard deviation of the results demonstrate that our method outperforms existing SOTA methods in both image quality and identity preservation.

\subsubsection{Ablation Study}

\noindent\textbf{Model Components.}
We conduct ablation studies on the AVSpeech dataset to validate the effectiveness of different components. 
The comparison results for different versions are listed in Table \ref{tab:ablation results on VoxCeleb}. 
It is evident that incorporating ConRe leads to improvements in accuracy for both gender and age attributes. This suggests that the identity information shared between the face and speech is effectively aligned and preserved through pre-training.  
With the addition of SAW, the feature distances between generated images and original portraits decrease, indicating that the synthesized results closely resemble the appearance of the original images. 
Furthermore, we provide visual examples in Fig. \ref{fig:ablation_cl} to illustrate the generated portrait images.  It is clear that with the inclusion of ConRe and SAW-FP, the generated face images exhibit a similar appearance and attributes to the speaker in the corresponding speech.

\noindent \textbf{Analysis of Statistical Face Prior.} 
To examine the effect of the number $N$ of face images used in computing the statistical face prior, we extract facial features from datasets of varying sizes: $N = 100, 500, 1000, 5000, 10000$, and $15000$. We then measure the feature differences between priors calculated at successive sample sizes using the L1 distance metric. Experimental results show that as $N$ increases, the statistical prior gradually stabilizes. In particular, the difference between $N = 10000$ and $N = 15000$ is minimal, indicating that the prior has effectively converged and is representative. The detailed trend is illustrated in Figure~\ref{fig:face_prior_convegence}.

To investigate the influence of gender composition on the statistical face prior, we conduct a controlled experiment by constructing five data subsets with varying gender ratios, each containing $N = 10000$ face images. The gender ratios are set to: $100\%$ female, $75\%$ female $+$ $25\%$ male, $50\%$ female $+$ $50\%$ male, $25\%$ female $+$ $75\%$ male, and $100\%$ male. We use the $50\%{:}50\%$ gender-balanced prior as the reference and compute the L1 distance between it and the priors derived from the other subsets.
As illustrated in Figure~\ref{fig:face_prior_gender}, the L1 distance increases as the gender ratio deviates from balance, indicating that the statistical face prior undergoes noticeable gender-specific shifts in facial morphological features. These findings underscore the importance of adopting a gender-balanced prior, which helps mitigate bias toward a particular gender and ensures better generalization across diverse speakers.

\begin{figure}
    \centering
    \includegraphics[width=0.85\linewidth]{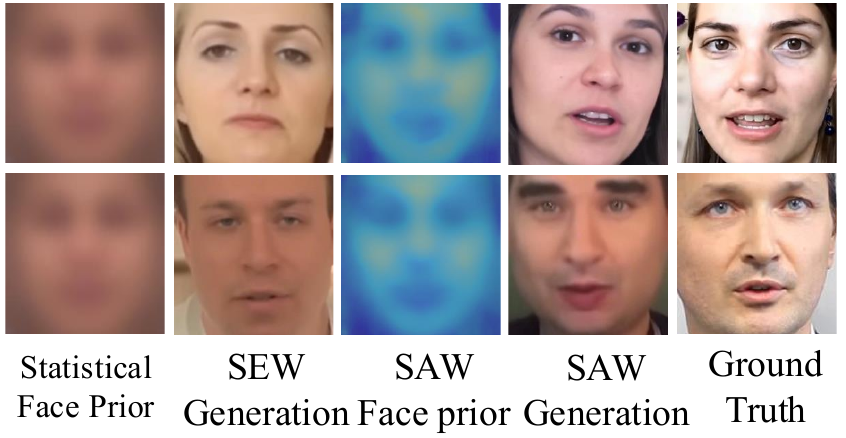}
    \caption{Visualization comparison of face prior weight.}
    \label{fig:saw_figure}
\end{figure}

\noindent \textbf{Sample Adaptive Weighted Mechanism.}
To investigate the effectiveness of sample adaptive weight, we perform experiments to evaluate the impact of varying the weight of the face prior $\beta$.
In Fig. \ref{fig:figure_prior} and Table \ref{tab:Ablation results on prior normalization weight.},
we present the comparison results between using a sample-equivalent weight $\beta^0 = 0.01 \times \mathbf{1}, \quad \mathbf{1} \in \mathbb{R}^{\text{dim}(z^p)}$, and a sample-adaptive weight $\beta$ calculated according to Eq. \ref{eq:weight}. It can be observed that utilizing the sample-adaptive weight for the face prior achieves better performance, highlighting the effectiveness of speaker discrimination through the introduction of a sample-adaptive weighted face prior. 
To further interpret this effect, we visualize the weight score generated by the SAW module over the statistical face prior, along with the final synthesized portraits in Figure \ref{fig:saw_figure}. It can be seen that assigns higher weights to facial priors that better match the speaker's identity, leading to more consistent and realistic facial details in the generated portraits. 

We further compare SAW against an attention-based variant to assess design efficiency. Table \ref{tab:ablation_structural_design} demonstrates that while attention yields marginal gains, it requires 4$\times$ more computation. This confirms SAW's optimal balance: preserving 99.7\% of attention's accuracy at 25\% computational cost.

\noindent \textbf{Effect of Speech Encoder.}
We evaluate the impact of the speech encoder on the Speech-to-Portrait (S2P) module. Specifically, we compare a CNN-based speech encoder trained from scratch with a pre-trained Wav2Vec model \cite{baevski2020wav2vec}. As shown in Table \ref{tab:speech_encoder_S2P}, the Wav2Vec-based model yields better identity preservation and facial appearance quality. We attribute this to its large-scale pretraining, which allows it to capture richer speaker-specific representations essential for accurate portrait synthesis. In future work, we plan to explore stronger pretrained speech encoders to further improve identity fidelity and overall generation quality.

\begin{table*}[htbp]
\centering
\caption{: Quantitative comparisons with previous speech-driven talking face generation methods on wild dataset. The best results are highlighted in bold.  }
\begin{tabular}{l|c|cc|cccc|c}
\hline
\multirow{2}{*}{Method} &\multirow{2}{*}{Venue} & \multicolumn{2}{c|}{Lip Consistency} & \multicolumn{4}{c|}{Visual Quality} & Temporal Consistency\\
\cline{3-9}
   & & LSE-D $\downarrow$& LSE-C $\uparrow$ & LPIPS $\downarrow$ & SSIM $\uparrow$& PSNR $\uparrow$ & FID $\downarrow$ & MAD $\downarrow$ \\
\hline
Aniportait \cite{wei2024aniportrait} &ECCV 2024&14.64 &4.42 &0.45 &0.63 &28.14 & 54.85 & 0.0045\\

Real3D-portrait \cite{ye2024real3d} & ICLR 2024&19.07 &3.27&0.56& 0.64&28.47& 48.38& 0.0056\\

SyncTalk \cite{peng2024synctalk} &CVPR 2024  &13.55 &4.55&0.39 & 0.68&28.82&56.47& 0.0063\\

\hline

Hallo \cite{xu2024hallo} &arXiv 2024.7 & 18.57&5.90&0.33&0.64 &28.67 & 46.59&0.0062\\

Anitalker \cite{liu2024anitalker} &ACM MM 2024&19.50 &4.26 &0.36&0.68& 28.73& 48.96& 0.0041\\
\hline
VideoRetalking \cite{cheng2022videoretalking} &SIGGRAPH 2022&12.56 &4.21 &0.17&0.63& 29.51 & 41.19& 0.0049\\
\hline
HRTF (Ours)&- & \textbf{8.89}&\textbf{9.23} &\textbf{0.14} &\textbf{0.75}& \textbf{29.68} &\textbf{28.76}& \textbf{0.0035} \\
HRTF-SCFP (Ours)&- & 9.94&8.68&0.18&0.73& 29.21&  29.19& 0.0038\\
\hline
\end{tabular}
\label{tab:avs-tf}
\end{table*}

 \begin{figure*}[htbp]
    \centering
    \includegraphics[width=0.99\linewidth]{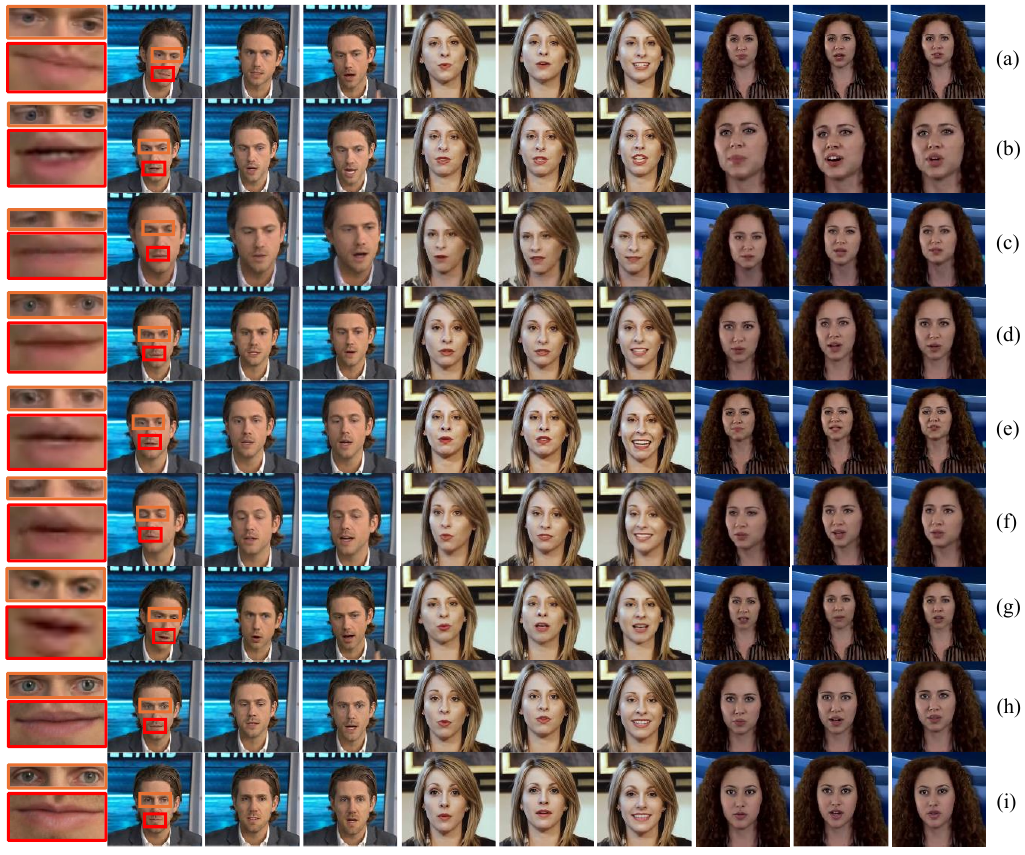}
    \caption{We present a qualitative comparison with existing approaches in speech-driven talking face generation. The first sample is selected from VoxCeleb, the second from HDTF, and the third from the wild dataset collected from AVSpeech. (a) Ground truth; (b) Generation by AniPortrait; (c) Generation by Real3D-portrait; (d) Generation by SyncTalk; (e) Generation by Hallo; (f) Generation by AniTalker; (g) Generation by VideoRetalking;
    (h) Generation by HDTF (Ours); (i) Generation by HDTF-SCFP (Ours). The areas inside the orange and red bounding boxes highlight the zoom-in details of the eyes and lips, respectively.}
    \label{fig:tf_sota}
\end{figure*}

 \subsection{Talking Face Generation}
\subsubsection{Comparison with SOTA methods.}
We compare our \textbf{HRTF} approach with five existing methods, categorized as follows: (1) Intermediate Representation-Based Methods: AniPortrait~\cite{wei2024aniportrait}, SyncTalk~\cite{peng2024synctalk}, and Real3D-Portrait~\cite{ye2024real3d}, which utilize intermediate representations to guide video generation.
(2) Latent Motion Representation-Based Methods: Hallo~\cite{xu2024hallo} and AniTalker~\cite{liu2024anitalker}, which encode holistic motion features in the latent space for audio-driven video synthesis.
(3) Speech-driven Lip Editing Method:
VideoRetalking \cite{cheng2022videoretalking}.

Specifically, we compare two configurations of our method:  
(1) \textbf{HRTF}: Uses the original reference images to synthesize talking head videos, allowing us to evaluate the performance of the talking head generation pipeline.
(2) \textbf{HRTF-SCFP}: Uses generated portraits as reference images for talking head generation, enabling the evaluation of the overall framework.

\noindent\textbf{Quantitative Comparison.}
The quantitative results on the VoxCeleb, HDTF, and self-collected wild datasets are presented in Table~\ref{tab:Vox-tf}, Table~\ref{tab:hdtf-tf}, and Table~\ref{tab:avs-tf}, respectively. The proposed HRTF talking face generation framework outperforms SOTA methods across all evaluated metrics. Specifically, the LSE-D score of our method achieves the highest performance, highlighting the accuracy of our lip-sync generation. Furthermore, our method achieves the best SSIM score, demonstrating that the generated video frames closely resemble the ground truth in terms of facial expressions, appearance, and head movement. Our HRTF-SCFP also delivers strong results, further validating the identity consistency and effectiveness of the proposed prior-guided speech-to-portrait generation pipeline.

\noindent\textbf{Qualitative Comparison.}
The qualitative comparison shown in Figure~\ref{fig:tf_sota} demonstrates the effectiveness of our approach in generating high-resolution video frames that align well with the given speech. Compared to latent motion representation-based methods such as Hallo~\cite{xu2024hallo} and AniTalker~\cite{liu2024anitalker}, our HRTF method exhibits superior lip consistency, which can be attributed to the effective integration of lip refinement within the holistic motion representation. Furthermore, the use of a high-resolution codebook enhances the visual quality of the generated frames. Upon examining the zoomed-in details, it is evident that the compared methods struggle to synthesize fine-grained facial textures and lip movements that are consistent with the speech. In contrast, our method excels in generating high-quality images, even when the source image is blurred.

\subsubsection{Ablation Study} 
\begin{figure}[ht]
    \centering
    \includegraphics[width=0.99\linewidth]{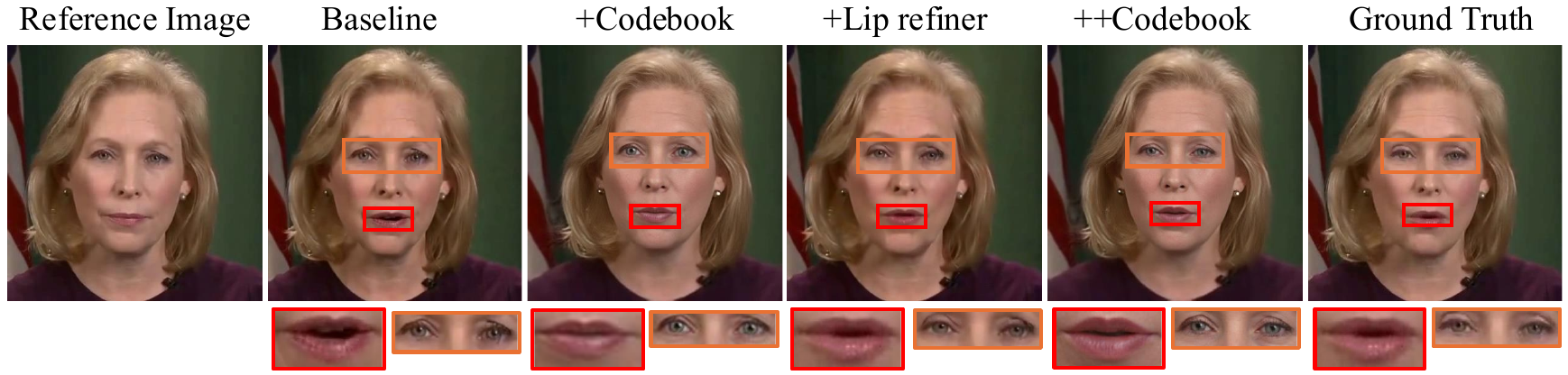}
    \caption{Visualization results of ablation study. The areas inside the orange and red bounding boxes highlight the zoom-in details of the eyes and lips, respectively.}
    \label{fig:ab}
\end{figure}

\textbf{Motion components.}
We conduct an ablation study on HRTF with the following variants:
Starting with a baseline model, which includes the speech encoder, image encoder and decoder, identity encoder, motion encoder, wrapping network, and motion diffusion, we progressively add the lip refiner and high-resolution codebook to assess the impact of each module. The results, shown in Fig.~\ref{fig:ab} and Table~\ref{tab:ab}, reveal that the baseline model, while capable of wrapping the reference image, lacks detailed lip movement information. The addition of the lip refiner effectively integrates lip motion into the holistic motion representation, resulting in improved lip enhancement. Finally, the inclusion of the high-resolution codebook further enhances the image fidelity and clarity, underscoring the contributions of each component to the overall performance.

\noindent \textbf{Effect of Speech Encoder.}
To assess the impact of different speech encoders in speech-driven talking head generation, we conduct comparative experiments between our adopted HuBERT model \cite{hsu2021hubert} and Wav2Vec \cite{baevski2020wav2vec}. As shown in Table~\ref{tab:speech_encoder_talkinghead}, the HuBERT-based model achieves superior lip synchronization and overall motion expressiveness. We attribute this improvement to HuBERT's ability to capture phoneme-level semantic representations through self-supervised clustering, which better aligns with the temporal structure of speech. This fine-grained modeling is particularly advantageous for synthesizing nuanced facial movements, such as lip articulation and subtle expressions.

\begin{table*}[htbp]
\centering
\caption{Ablation study on VoxCeleb dataset. The best results are highlighted in bold.  }
\begin{tabular}{ccc|cc|cccc|c}
\hline
  \multicolumn{3}{c|}{Method} & \multicolumn{2}{c|}{Lip Consistency} & \multicolumn{4}{c|}{Visual Quality}& Temporal Consistency \\
\hline
  Baseline&Lip refiner& Codebook & LSE-D $\downarrow$& LSE-C $\uparrow$ & LPIPS $\downarrow$ & SSIM $\uparrow$& PSNR $\uparrow$ & FID $\downarrow$ & MAD $\downarrow$\\
\hline
   $\checkmark$&&& 8.38&6.95 &0.15&0.63& 29.21&36.78 & 0.0058\\
    $\checkmark$&&$\checkmark$& 8.24&7.13 &0.13&0.64& 29.64&29.31& 0.0048 \\

 $\checkmark$&$\checkmark$&& 7.10 & 8.44&0.12&0.66 & 29.62&33.06& 0.0054 \\

$\checkmark$&$\checkmark$&$\checkmark$& \textbf{6.61}&\textbf{8.65} &\textbf{0.11} &\textbf{0.67}& \textbf{29.66} &\textbf{29.28}& \textbf{0.0042} \\
\hline
\end{tabular}
\label{tab:ab}
\end{table*}

\begin{table*}[t]
\centering
\caption{Ablation study on the impact of different speech encoders in speech-driven talking head generation. The best results are highlighted in bold.}
\label{tab:speech_encoder_talkinghead}
\begin{tabular}{lcc|cccc|c}
\hline
\multirow{2}{*}{Method} & \multicolumn{2}{c|}{Lip Consistency} & \multicolumn{4}{c|}{Visual Quality}&Temporal Consistency \\
\cline{2-8}
& LSE-D $\downarrow$ & LSE-C $\uparrow$ & LPIPS $\downarrow$ & SSIM $\uparrow$ &PSNR $\uparrow$ & FID $\downarrow$ & MAD $\downarrow$\\
\hline
Wav2Vec & 7.46 & 7.78 & 0.13 & 0.61 & 29.57 & 31.24 & 0.0045\\
Ours     & \textbf{6.61} & \textbf{8.65} & \textbf{0.11} & \textbf{0.67} & \textbf{29.66} & \textbf{29.28} & \textbf{0.0042}\\
\hline
\end{tabular}
\end{table*}

\begin{table}[t]
\centering
\caption{Comparison of total parameters (params), GPU memory usage, and inference speed of SOTA talking-head generation methods on a single A6000 GPU.}
\label{tab:efficiency}
\begin{tabular}{lccc}
\toprule
\textbf{Method} & \textbf{Params (M)} & \textbf{Memory (MB)} & \textbf{Speed (FPS)} \\
\midrule
Aniportrait~\cite{wei2024aniportrait}      & 2901 & 18551 & 1.00 \\
Real3D-Portrait~\cite{ye2024real3d} & 343  & 5186  & 0.74 \\
SyncTalk~\cite{peng2024synctalk}           & 109  & 808   & 2.03 \\
Hallo~ \cite{xu2024hallo}                  & 2519 & 8153  & 0.28 \\
AniTalker~\cite{liu2024anitalker}          & 421 & 3372  & 7.36 \\
VideoRetalking \cite{cheng2022videoretalking}& 299& 3092& 0.42\\
\hline
HDTF (ours)               & 470  & 3694  & 5.46 \\
HDTF-SCFP (ours)           & 1110 & 3694  & 5.46 \\
\bottomrule
\end{tabular}
\end{table}

\subsubsection{Efficiency Analysis} 
We evaluate the computational efficiency of our framework. The Speech-to-Portrait module generates a 256×256 facial image in approximately 2.3 seconds, while the Talking-Head Generation module synthesizes a 5-second video in around 22.8 seconds on a single A6000 GPU.
To further validate efficiency, we compare our method with SOTA approaches in terms of total parameters, GPU memory usage, and inference time (see Table~\ref{tab:efficiency}). Despite the two-stage framework, our method achieves comparable computational efficiency. As the two modules in our framework are executed sequentially, we report the peak GPU memory usage across both stages.

\begin{figure}[htbp]
    \centering
    \includegraphics[width=0.99\linewidth]{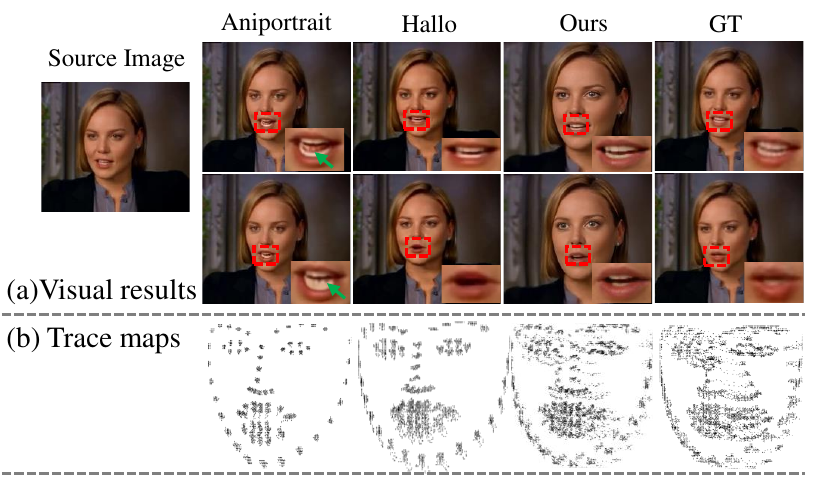}
    \caption{
    Further Analysis of Motion Representation. (a) Visualization of generated video frames: We compare our HRTF with AniPortrait~\cite{wei2024aniportrait} and Hallo~\cite{xu2024hallo} for talking face generation. The results demonstrate that our approach achieves superior expression modeling. (b) Trace maps of the generated frames: We visualize the trace maps of facial landmarks from the generated videos, showcasing the motion diversity achieved by our method. Please zoom in for more details.} 
    \label{fig:motion_case}
\end{figure}

\subsubsection{Case study}

To further validate the effectiveness of our method in addressing the aforementioned challenges in talking face generation, we present a case study comparing our approach with the intermediate representation-based method AniPortrait~\cite{wei2024aniportrait} and the latent motion representation-based method Hallo~\cite{xu2024hallo}, focusing on a sample with visible teeth (Fig.~\ref{fig:motion_case}). 
Intermediate representation-based methods like AniPortrait~\cite{wei2024aniportrait} rely heavily on the reference image to construct intermediate representations. As shown in Fig.~\ref{fig:motion_case}(a), this reliance causes interference from the mouth shape in the reference image, leading to inaccurate generation results. Furthermore, as illustrated by the trace map in Fig.~\ref{fig:motion_case}(b), AniPortrait primarily models lip movements, resulting in unrealistic outputs with limited expression dynamics. 
Latent motion representation-based methods like Hallo~\cite{xu2024hallo} demonstrate limited motion diversity in their generated results and struggle to accurately capture variations in lip and teeth movements. 
In contrast, our method effectively synthesizes realistic teeth while maintaining high audio-lip synchronization, leveraging the advantages of holistic motion representation, a lip refinement module, and a high-resolution decoder. As evidenced by the trace map in Fig.~\ref{fig:motion_case}(b), our approach achieves motion diversity comparable to the ground truth, demonstrating its effectiveness in overcoming these challenges.

\section{Conclusion}
\label{sec:conclusion}
In this work, we present a novel system capable of generating high-resolution talking faces with natural expressions from a single audio input, effectively addressing the key challenges in this domain. Our framework consists of two stages: SCFP, which estimates a high-quality speaker's face portrait with identity consistency guided by a statistical face prior, and HRTF, which synthesizes talking video frames featuring expressive dynamics such as lip movements and facial expressions. A region enhancement module further refines lip motion consistency, while a transformer-based codebook enhances video resolution. Extensive experiments on the HDTF, VoxCeleb, and AVSpeech datasets validate the effectiveness of our approach, which, to the best of our knowledge, is the first to achieve high-resolution, high-quality talking face generation using only audio input.

\section{Acknowledgments}
This work was supported by the National Natural Science Foundation of China (No. 62471420), GuangDong Basic and Applied Basic Research Foundation (2025A1515012296), and CCF-Tencent Rhino-Bird Open Research Fund.

\bibliography{refer}

\begin{thebibliography}{10}
\providecommand{\url}[1]{#1}
\csname url@samestyle\endcsname
\providecommand{\newblock}{\relax}
\providecommand{\bibinfo}[2]{#2}
\providecommand{\BIBentrySTDinterwordspacing}{\spaceskip=0pt\relax}
\providecommand{\BIBentryALTinterwordstretchfactor}{4}
\providecommand{\BIBentryALTinterwordspacing}{\spaceskip=\fontdimen2\font plus
\BIBentryALTinterwordstretchfactor\fontdimen3\font minus \fontdimen4\font\relax}
\providecommand{\BIBforeignlanguage}[2]{{%
\expandafter\ifx\csname l@#1\endcsname\relax
\typeout{** WARNING: IEEEtran.bst: No hyphenation pattern has been}%
\typeout{** loaded for the language `#1'. Using the pattern for}%
\typeout{** the default language instead.}%
\else
\language=\csname l@#1\endcsname
\fi
#2}}
\providecommand{\BIBdecl}{\relax}
\BIBdecl

\bibitem{ye2024mimictalk}
Z.~Ye, T.~Zhong, Y.~Ren, Z.~Jiang, J.~Huang, R.~Huang, J.~Liu, J.~He, C.~Zhang, Z.~Wang \emph{et~al.}, ``Mimictalk: Mimicking a personalized and expressive 3d talking face in minutes,'' in \emph{The Thirty-eighth Annual Conference on Neural Information Processing Systems}.

\bibitem{ye2024real3d}
Z.~Ye, T.~Zhong, Y.~Ren, J.~Yang, W.~Li, J.~Huang, Z.~Jiang, J.~He, R.~Huang, J.~Liu \emph{et~al.}, ``Real3d-portrait: One-shot realistic 3d talking portrait synthesis,'' in \emph{The Twelfth International Conference on Learning Representations}.

\bibitem{wang2024high}
M.~Wang, S.~Zhao, X.~Dong, and J.~Shen, ``High-fidelity and high-efficiency talking portrait synthesis with detail-aware neural radiance fields,'' \emph{IEEE Transactions on Visualization \& Computer Graphics}, 2024.

\bibitem{liu2024anitalker}
T.~Liu, F.~Chen, S.~Fan, C.~Du, Q.~Chen, X.~Chen, and K.~Yu, ``Anitalker: animate vivid and diverse talking faces through identity-decoupled facial motion encoding,'' in \emph{Proceedings of the 32nd ACM International Conference on Multimedia}, 2024, pp. 6696--6705.

\bibitem{xu2024hallo}
M.~Xu, H.~Li, Q.~Su, H.~Shang, L.~Zhang, C.~Liu, J.~Wang, Y.~Yao, and S.~Zhu, ``Hallo: Hierarchical audio-driven visual synthesis for portrait image animation,'' \emph{preprint arXiv:2406.08801}, 2024.

\bibitem{xu2024facechain}
C.~Xu, Y.~Liu, J.~Xing, W.~Wang, M.~Sun, J.~Dan, T.~Huang, S.~Li, Z.-Q. Cheng, Y.~Tai \emph{et~al.}, ``Facechain-imagineid: Freely crafting high-fidelity diverse talking faces from disentangled audio,'' in \emph{Proceedings of the IEEE/CVF Conference on Computer Vision and Pattern Recognition}, 2024, pp. 1292--1302.

\bibitem{wei2024aniportrait}
H.~Wei, Z.~Yang, and Z.~Wang, ``Aniportrait: Audio-driven synthesis of photorealistic portrait animations,'' \emph{preprint arXiv:2403.17694}, 2024.

\bibitem{zhong2023identity}
W.~Zhong, C.~Fang, Y.~Cai, P.~Wei, G.~Zhao, L.~Lin, and G.~Li, ``Identity-preserving talking face generation with landmark and appearance priors,'' in \emph{IEEE/CVF Conference on Computer Vision and Pattern Recognition}, 2023, pp. 9729--9738.

\bibitem{ma2023dreamtalk}
Y.~Ma, S.~Zhang, J.~Wang, X.~Wang, Y.~Zhang, and Z.~Deng, ``Dreamtalk: When expressive talking head generation meets diffusion probabilistic models,'' \emph{preprint arXiv:2312.09767}, 2023.

\bibitem{danvevcek2022emoca}
R.~Dan{\v{e}}{\v{c}}ek, M.~J. Black, and T.~Bolkart, ``Emoca: Emotion driven monocular face capture and animation,'' in \emph{Proceedings of the IEEE/CVF Conference on Computer Vision and Pattern Recognition}, 2022, pp. 20\,311--20\,322.

\bibitem{peng2024synctalk}
Z.~Peng, W.~Hu, Y.~Shi, X.~Zhu, X.~Zhang, H.~Zhao, J.~He, H.~Liu, and Z.~Fan, ``Synctalk: The devil is in the synchronization for talking head synthesis,'' in \emph{Proceedings of the IEEE/CVF Conference on Computer Vision and Pattern Recognition}, 2024, pp. 666--676.

\bibitem{chen2023improving}
Q.~Chen, Z.~Ma, T.~Liu, X.~Tan, Q.~Lu, K.~Yu, and X.~Chen, ``Improving few-shot learning for talking face system with tts data augmentation,'' in \emph{ICASSP 2023-2023 IEEE International Conference on Acoustics, Speech and Signal Processing (ICASSP)}.\hskip 1em plus 0.5em minus 0.4em\relax IEEE, 2023, pp. 1--5.

\bibitem{peng2023emotalk}
Z.~Peng, H.~Wu, Z.~Song, H.~Xu, X.~Zhu, J.~He, H.~Liu, and Z.~Fan, ``Emotalk: Speech-driven emotional disentanglement for 3d face animation,'' in \emph{Proceedings of the IEEE/CVF International Conference on Computer Vision}, 2023, pp. 20\,687--20\,697.

\bibitem{xu2024vasa}
S.~Xu, G.~Chen, Y.-X. Guo, J.~Yang, C.~Li, Z.~Zang, Y.~Zhang, X.~Tong, and B.~Guo, ``Vasa-1: Lifelike audio-driven talking faces generated in real time,'' \emph{preprint arXiv:2404.10667}, 2024.

\bibitem{ho2022cascaded}
J.~Ho, C.~Saharia, W.~Chan, D.~J. Fleet, M.~Norouzi, and T.~Salimans, ``Cascaded diffusion models for high fidelity image generation,'' \emph{Journal of Machine Learning Research}, vol.~23, no.~47, pp. 1--33, 2022.

\bibitem{ling2024posetalk}
J.~Ling, Y.~Wang, H.~Xue, R.~Xie, and L.~Song, ``Posetalk: Text-and-audio-based pose control and motion refinement for one-shot talking head generation,'' \emph{preprint arXiv:2409.02657}, 2024.

\bibitem{bull1983voice}
R.~Bull, H.~Rathborn, and B.~R. Clifford, ``The voice-recognition accuracy of blind listeners,'' \emph{Perception}, vol.~12, no.~2, pp. 223--226, 1983.

\bibitem{deaton2010understanding}
A.~Deaton, ``Understanding the mechanisms of economic development,'' \emph{Journal of Economic Perspectives}, vol.~24, no.~3, pp. 3--16, 2010.

\bibitem{schweinberger2014speaker}
S.~R. Schweinberger, H.~Kawahara, A.~P. Simpson, V.~G. Skuk, and R.~Z{\"a}ske, ``Speaker perception,'' \emph{Wiley Interdisciplinary Reviews: Cognitive Science}, vol.~5, no.~1, pp. 15--25, 2014.

\bibitem{razavi2019generating}
A.~Razavi, A.~Van~den Oord, and O.~Vinyals, ``Generating diverse high-fidelity images with vq-vae-2,'' \emph{Advances in neural information processing systems}, vol.~32, 2019.

\bibitem{zhou2022towards}
S.~Zhou, K.~Chan, C.~Li, and C.~C. Loy, ``Towards robust blind face restoration with codebook lookup transformer,'' \emph{Advances in Neural Information Processing Systems}, vol.~35, pp. 30\,599--30\,611, 2022.

\bibitem{zhang2021flow}
Z.~Zhang, L.~Li, Y.~Ding, and C.~Fan, ``Flow-guided one-shot talking face generation with a high-resolution audio-visual dataset,'' in \emph{Proceedings of the IEEE/CVF Conference on Computer Vision and Pattern Recognition}, 2021, pp. 3661--3670.

\bibitem{chung2018voxceleb2}
J.~S. Chung, A.~Nagrani, and A.~Zisserman, ``Voxceleb2: Deep speaker recognition,'' \emph{preprint arXiv:1806.05622}, 2018.

\bibitem{ephrat2018looking}
A.~Ephrat, I.~Mosseri, O.~Lang, T.~Dekel, K.~Wilson, A.~Hassidim, W.~T. Freeman, and M.~Rubinstein, ``Looking to listen at the cocktail party: A speaker-independent audio-visual model for speech separation,'' \emph{preprint arXiv:1804.03619}, 2018.

\bibitem{duarte2019wav2pix}
A.~Duarte, F.~Roldan, M.~Tubau, J.~Escur, S.~Pascual, A.~Salvador, E.~Mohedano, K.~McGuinness, J.~Torres, and X.~Giro-i Nieto, ``Wav2pix: Speech-conditioned face generation using generative adversarial networks,'' in \emph{ICASSP 2019 IEEE International Conference on Acoustics, Speech and Signal Processing}, 2019, pp. 8633--8637.

\bibitem{wen2019face}
Y.~Wen, R.~Singh, and B.~Raj, ``Face reconstruction from voice using generative adversarial networks,'' in \emph{the 33rd International Conference on Neural Information Processing Systems}, 2019, pp. 5265--5274.

\bibitem{fang2022facial}
Z.~Fang, Z.~Liu, T.~Liu, C.-C. Hung, J.~Xiao, and G.~Feng, ``Facial expression gan for voice-driven face generation,'' \emph{The Visual Computer}, vol.~38, no.~3, pp. 1151--1164, 2022.

\bibitem{choi2019inference}
H.-S. Choi, C.~Park, and K.~Lee, ``From inference to generation: End-to-end fully self-supervised generation of human face from speech,'' in \emph{International Conference on Learning Representations}, 2019.

\bibitem{dhariwal2021diffusion}
P.~Dhariwal and A.~Nichol, ``Diffusion models beat gans on image synthesis,'' \emph{Advances in neural information processing systems}, vol.~34, pp. 8780--8794, 2021.

\bibitem{oh2019speech2face}
T.-H. Oh, T.~Dekel, C.~Kim, I.~Mosseri, W.~T. Freeman, M.~Rubinstein, and W.~Matusik, ``Speech2face: Learning the face behind a voice,'' in \emph{IEEE/CVF Conference on Computer Vision and Pattern Recognition}, 2019, pp. 7539--7548.

\bibitem{bai2022speech}
Y.~Bai, T.~Ma, L.~Wang, and Z.~Zhang, ``Speech fusion to face: Bridging the gap between human's vocal characteristics and facial imaging,'' in \emph{30th ACM International Conference on Multimedia}, 2022, pp. 2042--2050.

\bibitem{sun2023vividtalk}
X.~Sun, L.~Zhang, H.~Zhu, P.~Zhang, B.~Zhang, X.~Ji, K.~Zhou, D.~Gao, L.~Bo, and X.~Cao, ``Vividtalk: One-shot audio-driven talking head generation based on 3d hybrid prior,'' \emph{preprint arXiv:2312.01841}, 2023.

\bibitem{zhang2023sadtalker}
W.~Zhang, X.~Cun, X.~Wang, Y.~Zhang, X.~Shen, Y.~Guo, Y.~Shan, and F.~Wang, ``Sadtalker: Learning realistic 3d motion coefficients for stylized audio-driven single image talking face animation,'' in \emph{Proceedings of the IEEE/CVF Conference on Computer Vision and Pattern Recognition}, 2023, pp. 8652--8661.

\bibitem{2024real3d}
Z.~Ye, T.~Zhong, Y.~Ren, J.~Yang, W.~Li, J.~Huang, Z.~Jiang, J.~He, R.~Huang, J.~Liu \emph{et~al.}, ``Real3d-portrait: One-shot realistic 3d talking portrait synthesis,'' in \emph{The Twelfth International Conference on Learning Representations}.

\bibitem{tian2025emo}
L.~Tian, Q.~Wang, B.~Zhang, and L.~Bo, ``Emo: Emote portrait alive generating expressive portrait videos with audio2video diffusion model under weak conditions,'' in \emph{European Conference on Computer Vision}.\hskip 1em plus 0.5em minus 0.4em\relax Springer, 2025, pp. 244--260.

\bibitem{ho2022imagen}
J.~Ho, W.~Chan, C.~Saharia, J.~Whang, R.~Gao, A.~Gritsenko, D.~P. Kingma, B.~Poole, M.~Norouzi, D.~J. Fleet \emph{et~al.}, ``Imagen video: High definition video generation with diffusion models,'' \emph{preprint arXiv:2210.02303}, 2022.

\bibitem{podell2023sdxl}
D.~Podell, Z.~English, K.~Lacey, A.~Blattmann, T.~Dockhorn, J.~M{\"u}ller, J.~Penna, and R.~Rombach, ``Sdxl: Improving latent diffusion models for high-resolution image synthesis,'' \emph{preprint arXiv:2307.01952}, 2023.

\bibitem{blattmann2023align}
A.~Blattmann, R.~Rombach, H.~Ling, T.~Dockhorn, S.~W. Kim, S.~Fidler, and K.~Kreis, ``Align your latents: High-resolution video synthesis with latent diffusion models,'' in \emph{Proceedings of the IEEE/CVF Conference on Computer Vision and Pattern Recognition}, 2023, pp. 22\,563--22\,575.

\bibitem{skorokhodov2024hierarchical}
I.~Skorokhodov, W.~Menapace, A.~Siarohin, and S.~Tulyakov, ``Hierarchical patch diffusion models for high-resolution video generation,'' in \emph{Proceedings of the IEEE/CVF Conference on Computer Vision and Pattern Recognition}, 2024, pp. 7569--7579.

\bibitem{hu2022global}
M.~Hu, Y.~Wang, T.-J. Cham, J.~Yang, and P.~N. Suganthan, ``Global context with discrete diffusion in vector quantised modelling for image generation,'' in \emph{Proceedings of the IEEE/CVF Conference on Computer Vision and Pattern Recognition}, 2022, pp. 11\,502--11\,511.

\bibitem{yan2021videogpt}
W.~Yan, Y.~Zhang, P.~Abbeel, and A.~Srinivas, ``Videogpt: Video generation using vq-vae and transformers,'' \emph{preprint arXiv:2104.10157}, 2021.

\bibitem{chen2024stylespeech}
X.~Chen, X.~Wang, S.~Zhang, L.~He, Z.~Wu, X.~Wu, and H.~Meng, ``Stylespeech: Self-supervised style enhancing with vq-vae-based pre-training for expressive audiobook speech synthesis,'' in \emph{ICASSP 2024-2024 IEEE International Conference on Acoustics, Speech and Signal Processing (ICASSP)}.\hskip 1em plus 0.5em minus 0.4em\relax IEEE, 2024, pp. 12\,316--12\,320.

\bibitem{tan2024flowvqtalker}
S.~Tan, B.~Ji, and Y.~Pan, ``Flowvqtalker: High-quality emotional talking face generation through normalizing flow and quantization,'' in \emph{Proceedings of the IEEE/CVF Conference on Computer Vision and Pattern Recognition}, 2024, pp. 26\,317--26\,327.

\bibitem{katospeech}
S.~Kato and T.~Hashimoto, ``Speech-to-face conversion using denoising diffusion probabilistic models,'' in \emph{INTERSPEECH}, vol.~10, 2023, pp. 2023--1358.

\bibitem{choi2022snac}
B.-J. Choi, M.-J. Jeong, J.-Y. Lee, N.-S. Kim, and B.-J. Kim, ``Snac: Speaker-normalized affine coupling layer in flow-based architecture for zero-shot multi-speaker text-to-speech,'' \emph{IEEE Signal Processing Letters}, vol.~29, pp. 2502--2506, 2022.

\bibitem{yu2024gaussiantalker}
H.~Yu, Z.~Qu, Q.~Yu, J.~Chen, Z.~Jiang, Z.~Chen, S.~Zhang, J.~Xu, F.~Wu, C.~Lv \emph{et~al.}, ``Gaussiantalker: Speaker-specific talking head synthesis via 3d gaussian splatting,'' in \emph{Proceedings of the 32nd ACM International Conference on Multimedia}, 2024, pp. 3548--3557.

\bibitem{afham2022crosspoint}
M.~Afham, I.~Dissanayake, D.~Dissanayake, A.~Dharmasiri, K.~Thilakarathna, and R.~Rodrigo, ``Crosspoint: Self-supervised cross-modal contrastive learning for 3d point cloud understanding,'' in \emph{IEEE/CVF Conference on Computer Vision and Pattern Recognition}, 2022, pp. 9902--9912.

\bibitem{parelli2023clip}
M.~Parelli, A.~Delitzas, N.~Hars, G.~Vlassis, S.~Anagnostidis, G.~Bachmann, and T.~Hofmann, ``Clip-guided vision-language pre-training for question answering in 3d scenes,'' in \emph{IEEE/CVF Conference on Computer Vision and Pattern Recognition}, 2023, pp. 5606--5611.

\bibitem{radford2021learning}
A.~Radford, J.~W. Kim, C.~Hallacy, A.~Ramesh, G.~Goh, S.~Agarwal, G.~Sastry, A.~Askell, P.~Mishkin, J.~Clark \emph{et~al.}, ``Learning transferable visual models from natural language supervision,'' in \emph{International Conference on Machine Learning}, 2021, pp. 8748--8763.

\bibitem{qawaqneh2017deep}
Z.~Qawaqneh, A.~A. Mallouh, and B.~D. Barkana, ``Deep convolutional neural network for age estimation based on vgg-face model,'' \emph{preprint arXiv:1709.01664}, 2017.

\bibitem{woo2018cbam}
S.~Woo, J.~Park, J.-Y. Lee, and I.~S. Kweon, ``Cbam: Convolutional block attention module,'' in \emph{Proceedings of the European conference on computer vision (ECCV)}, 2018, pp. 3--19.

\bibitem{zhang2018unreasonable}
R.~Zhang, P.~Isola, A.~A. Efros, E.~Shechtman, and O.~Wang, ``The unreasonable effectiveness of deep features as a perceptual metric,'' in \emph{IEEE/CVF Conference on Computer Vision and Pattern Recognition}, 2018, pp. 586--595.

\bibitem{siarohin2019first}
A.~Siarohin, S.~Lathuili{\`e}re, S.~Tulyakov, E.~Ricci, and N.~Sebe, ``First order motion model for image animation,'' \emph{Advances in neural information processing systems}, vol.~32, 2019.

\bibitem{ni2023conditional}
H.~Ni, C.~Shi, K.~Li, S.~X. Huang, and M.~R. Min, ``Conditional image-to-video generation with latent flow diffusion models,'' in \emph{IEEE/CVF Conference on Computer Vision and Pattern Recognition}, 2023, pp. 18\,444--18\,455.

\bibitem{wang2024lia}
Y.~Wang, D.~Yang, F.~Bremond, and A.~Dantcheva, ``Lia: Latent image animator,'' \emph{IEEE Transactions on Pattern Analysis and Machine Intelligence}, 2024.

\bibitem{simonyan2014very}
K.~Simonyan, ``Very deep convolutional networks for large-scale image recognition,'' \emph{preprint arXiv:1409.1556}, 2014.

\bibitem{van2017neural}
A.~Van Den~Oord, O.~Vinyals \emph{et~al.}, ``Neural discrete representation learning,'' \emph{Advances in neural information processing systems}, vol.~30, 2017.

\bibitem{nagrani2017voxceleb}
A.~Nagrani, J.~S. Chung, and A.~Zisserman, ``Voxceleb: a large-scale speaker identification dataset,'' \emph{preprint arXiv:1706.08612}, 2017.

\bibitem{king2009dlib}
D.~E. King, ``Dlib-ml: A machine learning toolkit,'' \emph{The Journal of Machine Learning Research}, vol.~10, pp. 1755--1758, 2009.

\bibitem{rehman2012content}
M.~Rehman, M.~Iqbal, M.~Sharif, and M.~Raza, ``Content based image retrieval: survey,'' \emph{World Applied Sciences Journal}, vol.~19, no.~3, pp. 404--412, 2012.

\bibitem{wu2022retrievalguard}
Y.~Wu, H.~Zhang, and H.~Huang, ``Retrievalguard: Provably robust 1-nearest neighbor image retrieval,'' in \emph{International Conference on Machine Learning}.\hskip 1em plus 0.5em minus 0.4em\relax PMLR, 2022, pp. 24\,266--24\,279.

\bibitem{chung2017out}
J.~S. Chung and A.~Zisserman, ``Out of time: automated lip sync in the wild,'' in \emph{Computer Vision--ACCV 2016 Workshops: ACCV 2016 International Workshops, Taipei, Taiwan, November 20-24, 2016, Revised Selected Papers, Part II 13}.\hskip 1em plus 0.5em minus 0.4em\relax Springer, 2017, pp. 251--263.

\bibitem{chen2019hierarchical}
L.~Chen, R.~K. Maddox, Z.~Duan, and C.~Xu, ``Hierarchical cross-modal talking face generation with dynamic pixel-wise loss,'' in \emph{Proceedings of the IEEE/CVF conference on computer vision and pattern recognition}, 2019, pp. 7832--7841.

\bibitem{heusel2017gans}
M.~Heusel, H.~Ramsauer, T.~Unterthiner, B.~Nessler, and S.~Hochreiter, ``Gans trained by a two time-scale update rule converge to a local nash equilibrium,'' in \emph{Advances in Neural Information Processing Systems}, 2017, pp. 6629--6640.

\bibitem{szegedy2015going}
C.~Szegedy, W.~Liu, Y.~Jia, P.~Sermanet, S.~Reed, D.~Anguelov, D.~Erhan, V.~Vanhoucke, and A.~Rabinovich, ``Going deeper with convolutions,'' in \emph{Proceedings of the IEEE conference on computer vision and pattern recognition}, 2015, pp. 1--9.

\bibitem{wang2004image}
Z.~Wang, A.~C. Bovik, H.~R. Sheikh, and E.~P. Simoncelli, ``Image quality assessment: from error visibility to structural similarity,'' \emph{IEEE transactions on image processing}, vol.~13, no.~4, pp. 600--612, 2004.

\bibitem{teed2020raft}
Z.~Teed and J.~Deng, ``Raft: Recurrent all-pairs field transforms for optical flow,'' in \emph{European Conference on Computer Vision (ECCV)}.\hskip 1em plus 0.5em minus 0.4em\relax Springer International Publishing, 2020, pp. 402--419.

\bibitem{parkhi2015deep}
O.~Parkhi, A.~Vedaldi, and A.~Zisserman, ``Deep face recognition,'' in \emph{BMVC 2015-Proceedings of the British Machine Vision Conference 2015}.\hskip 1em plus 0.5em minus 0.4em\relax British Machine Vision Association, 2015.

\bibitem{rombach2022high}
R.~Rombach, A.~Blattmann, D.~Lorenz, P.~Esser, and B.~Ommer, ``High-resolution image synthesis with latent diffusion models,'' in \emph{IEEE/CVF Conference on Computer Vision and Pattern Recognition}, 2022, pp. 10\,684--10\,695.

\bibitem{hsu2021hubert}
W.-N. Hsu, B.~Bolte, Y.-H.~H. Tsai, K.~Lakhotia, R.~Salakhutdinov, and A.~Mohamed, ``Hubert: Self-supervised speech representation learning by masked prediction of hidden units,'' \emph{IEEE/ACM transactions on audio, speech, and language processing}, vol.~29, pp. 3451--3460, 2021.

\bibitem{karras2020analyzing}
T.~Karras, S.~Laine, M.~Aittala, J.~Hellsten, J.~Lehtinen, and T.~Aila, ``Analyzing and improving the image quality of stylegan,'' in \emph{Proceedings of the IEEE/CVF conference on computer vision and pattern recognition}, 2020, pp. 8110--8119.

\bibitem{cheng2022videoretalking}
K.~Cheng, X.~Cun, Y.~Zhang, M.~Xia, F.~Yin, M.~Zhu, X.~Wang, J.~Wang, and N.~Wang, ``Videoretalking: Audio-based lip synchronization for talking head video editing in the wild,'' 2022.

\bibitem{baevski2020wav2vec}
A.~Baevski, Y.~Zhou, A.~Mohamed, and M.~Auli, ``wav2vec 2.0: A framework for self-supervised learning of speech representations,'' in \emph{Advances in Neural Information Processing Systems}, vol.~33, 2020, pp. 12\,449--12\,460.

\end{thebibliography}
\bibliographystyle{IEEEtran}
\end{document}